\documentclass[
 reprint,
 amsmath,amssymb,
 aps,
 prb,
]{revtex4-2}
\usepackage{graphicx}
\usepackage{dcolumn}
\usepackage{bm}
\usepackage{subfigure}
\usepackage[dvipsnames]{xcolor}
\usepackage{amsmath}
\usepackage[normalem]{ulem}

\newcommand{\bra}[1]{\ensuremath{\langle#1|}}
\newcommand{\ket}[1]{\ensuremath{|{#1}\rangle}}

\usepackage[colorlinks]{hyperref}

\newcommand{\correction}[1]{#1}

\hypersetup{
    colorlinks = true,
    linkbordercolor = {white},
    linkcolor= BrickRed,
    citecolor=blue,	 
    }

\begin{document}

\title{\correction{Two-body Wigner molecularization in asymmetric quantum dot spin qubits}}
\author{Jos\'e C. Abadillo-Uriel}
\affiliation{Universit\'e Grenoble Alpes, CEA, IRIG-MEM-L\_SIM, 38000 Grenoble, France}%
\author{Biel Martinez}
\affiliation{Universit\'e Grenoble Alpes, CEA, IRIG-MEM-L\_SIM, 38000 Grenoble, France}%
\author{Michele Filippone}
\affiliation{Universit\'e Grenoble Alpes, CEA, IRIG-MEM-L\_SIM, 38000 Grenoble, France}%
\author{Yann-Michel Niquet}
\affiliation{Universit\'e Grenoble Alpes, CEA, IRIG-MEM-L\_SIM, 38000 Grenoble, France}%

\begin{abstract}
Coulomb interactions strongly influence the spectrum and the wave functions of few electrons or holes confined in a quantum dot. In particular, when the confinement potential is not too strong, the Coulomb repulsion triggers the formation of a correlated state, the Wigner molecule, where the particles tend to split apart. We show that the anisotropy of the confinement potential strongly enhances the molecularization process and affects the performances of quantum-dot systems used as spin qubits. Relying on analytical and numerical solutions of the two-particle problem -- both in a simplified single-band approximation and in realistic setups --  we highlight the exponential suppression of the singlet-triplet gap with increasing anisotropy. We compare the molecularization effects in different semiconductor materials and discuss how they specifically hamper Pauli spin blockade readout and reduce the exchange interactions in two-qubit gates.  
\end{abstract}

\maketitle

\section{Introduction}

In recent years, the performances of semiconductor quantum dot-based qubits \cite{Loss:1998p120,Zwanenburg:2013p961} have been steadily improving. Since the first demonstrations of single qubit gates \citep{Kim:2014p70, Veldhorst:2015p410, Maurand:2016p13575}, the fidelity of these operations has grown to the point of hitting the quantum error correction threshold \citep{YonedaNatNano2018, huang2019fidelity,cerfontaine2020closed}. While striving to reach this threshold consistently, the community has started to report successful experiments on two-qubit gates, also achieving high fidelities \citep{Nichol:2017p1, noiri2018fast, ZajacScience2018, WatsonNature2018, huang2019fidelity, XuePRX2019, Andrews:2019p05004, Hendrickx_Nature2020, Borjans:2020p195, petit2020universal, hendrickx2021four,xue2021computing}. 

Harnessing the intrinsic degrees of freedom~\citep{Bryant:1987p1140, Shaji:2008p540, Kodera:2009p22043, Chen:2017p035408} of each dot is paramount for the optimal operation of multi-qubit devices. Spin, valley and orbital degrees of freedom are thus currently exploited to define various kinds of qubits~\cite{DiVincenzo:2000p1642, Levy:2002p1446, Petta:2005p2180, Maune:2012p344, Shi:2012p140503, Koh:2013p19695, medford2013resonant, Taylor:2013p050502}, perform Pauli-spin blockade readout~\citep{Shaji:2008p540, Kodera:2009p22043, Simmons:2010p245312, Borselli:2011p063109, kotekar2017pauli} and exchange gates \citep{DiVincenzo:2000p1642, WatsonNature2018, hendrickx2021four,tariq2021impact}, or increase the qubit lifetimes~\citep{Leon:2020p797}. The design of the dots, which controls the interplay between these degrees of freedom, is therefore becoming a critical concern, especially in the prospect of upscaling multi-qubit systems. Interestingly, some architectures, such as dots confined along a one-dimensional (nanowire) channel \citep{VoisinNanoLett2014,BetzNL2015, Maurand:2016p13575, watzinger2018germanium}, can produce very anisotropic qubits. This anisotropy may be leveraged to enhance the qubit performances by, e.g., increasing the Rabi frequencies (as proposed in hole spin qubits~\cite{MichalPRB2021,bosco2021squeezed,bosco2021fully}).

\begin{figure}[t]
\centering
  \includegraphics[width=\columnwidth]{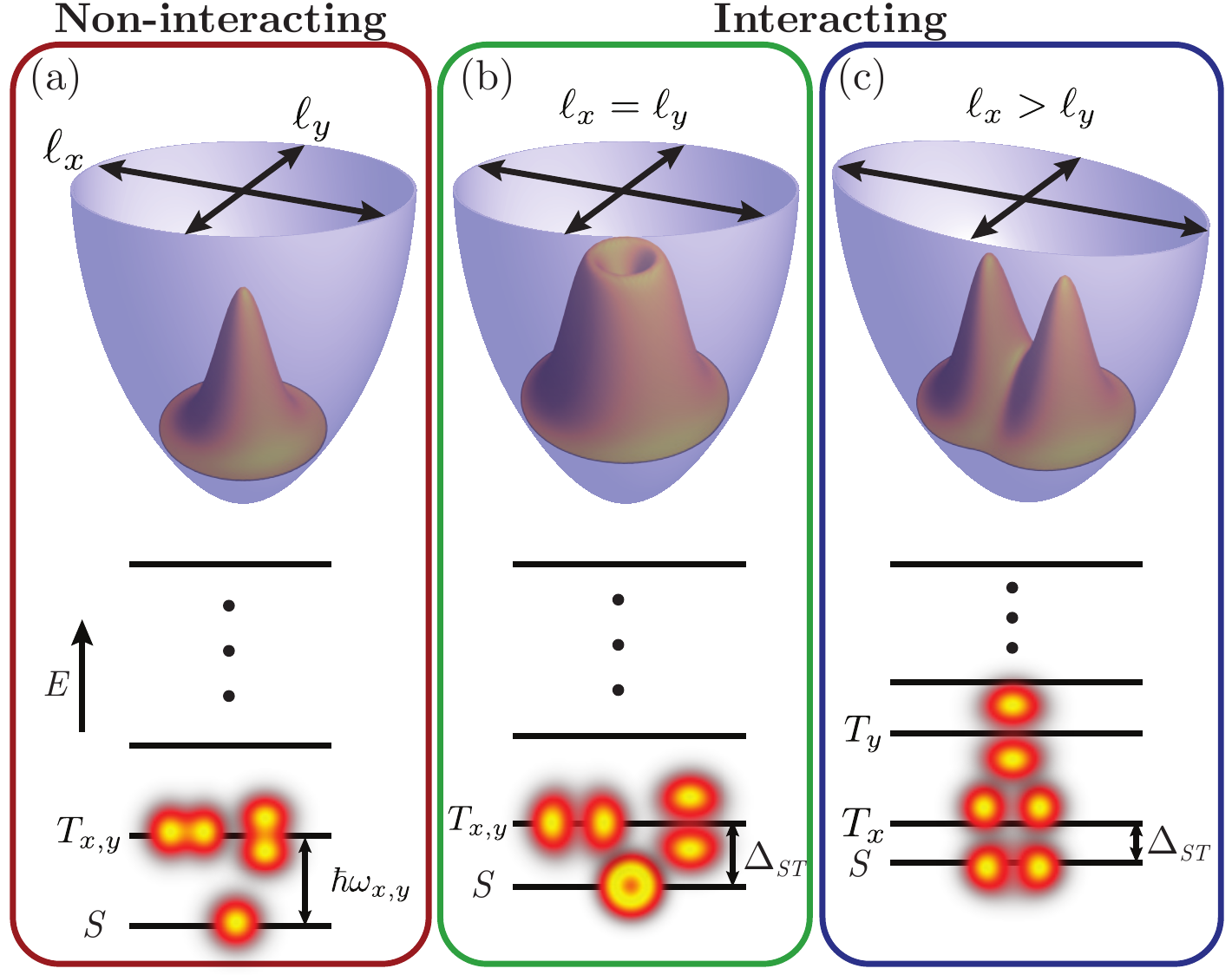}
\caption{Wigner molecularization controlled by confinement anisotropy in quantum dots. The panels represent the ground-state density for two particles confined in 2D harmonic potentials (top) and the associated excitation spectrum and wave function densities (bottom). In the non-interacting case (a), the singlet ground state is $s$-symmetric and the excited triplet state is obtained by promoting one particle to the $p_x$ (or $p_y$) orbital. In the interacting case (b), the ground state develops a dip at the center of the dot, accompanied by a closure of the singlet-triplet gap $\Delta_{ST}$, which is dramatically suppressed in the asymmetric case (c), where electrons/holes further separate along the weak confinement direction.}
\label{fig:schematics}
\end{figure}

Moreover, Coulomb interactions among electrons/holes in the quantum dots may also lead to outstanding many-body effects that highly depend on the system geometry. For instance, it is well established that the dimensionality of an electron gas is critical for the formation of Wigner crystals~\cite{Wigner:1934,Tanatar:1989}, as electrons have less freedom to avoid each other and minimize the Coulomb repulsion in fewer dimensions. In the context of few electrons or holes confined in quantum dots, the phenomenon of Wigner crystallization translates into {\it Wigner molecularization}~\citep{Bryant:1987p1140, Tanatar:1989, Egger:1999p3320, Yannouleas:1999p5325, Reimann:2000p8108,reimann2002electronic, reusch:PhysRevB.63.113313,Filinov:2001p3851,kalliakos2008molecular,Kristinsdottir:2011p041101, Pecker:2013p576}: in doubly-occupied dots, Coulomb interactions increase the cost of putting two electrons in the same $s$-like ground state, in a spin-singlet ($S$) configuration (see Fig.~\ref{fig:schematics}). As a consequence, the two electrons split apart, which reduces the gap $\Delta_{ST}$ with the spin triplet ($T$)~\citep{Ellenberger:2006p126806, Singha:2010p246802}.

This consideration explains the experimental evidence that the measured singlet-triplet gaps are generally smaller than the orbital splittings \correction{in materials without valleys \citep{Kouwenhoven:2001p701, Hanson:2007p1217}. In silicon devices, the valley splittings are more robust against Coulomb interactions \citep{ercan2021charge}, but an overall compression of the energy spectrum has been evidenced in large Si/SiO$_2$ and Si/SiGe quantum dots~\citep{lundberg2020spin, corrigan2020coherent} by the unexpected presence of multiple energy transitions in an small frequency range}. The singlet-triplet gap is a particularly relevant quantity for the operation of spin qubits, as it enables Pauli spin blockade (used for readout)~\citep{BetzNL2015, Chen:2017p035408} and exchange interactions (in two-qubit gates and singlet-triplet qubits)~\citep{Levy:2002p1446}. From the theoretical perspective, these effects have been studied in 2D quantum dots \citep{MerktPRB1991, Hada:2003p155322, Liu:2012p45311, Miserev:2019p205129}, with a particular attention to valley-orbit coupling for electrons in Si~\citep{ercan2021strong, ercan2021charge}.

In this work, we assess the dramatic enhancement of the molecularization effects caused by the interplay between Coulomb interactions and the anisotropic confinement in realistic quantum dot devices that are currently investigated as possible qubit platforms. We first consider a simplified single band model for two electrons/holes trapped in a harmonic potential. Relying on analytical results and an efficient numerical solution of the two-body problem, we provide a comprehensive description of the molecularization process. In particular, we characterize the spatial rearrangements of the ground and first excited states by Coulomb interactions and show the exponential suppression of the singlet-triplet gap $\Delta_{ST}$ by anisotropic confinement. We then discuss the role of materials in Wigner molecularization, and show that it is all the more prominent as the carrier masses are heavy. We finally strengthen our analysis by addressing realistic devices in silicon. \correction{For the sake of illustration, we focus on holes, which have no valley degree of freedom, and address the role of valleys for electrons in the appendices}. The single-hole states are first computed with a four bands $\mathbf{k}\cdot\mathbf{p}$ model, then are passed to a full configuration interaction (CI) solver for the interacting wave functions. This approach, while more expensive, is also more exhaustive, as it accounts for realistic geometries and electrostatics, as well as for the heavy-hole/light-hole mixing. It shows that Wigner molecularization is a concern in present qubit architectures based on one-dimensional channels such as nanowires. We discuss the implications for qubit readout and exchange interactions, showing that Coulomb interactions must be taken into account to achieve the most effective design of compact and high-performance multi-qubit platforms.

The paper is structured as follows: in Section~\ref{sec:model}, we introduce the parabolic single band model and describe the molecularization process based on analytical perturbative calculations complemented by the effective numerical solution of the two-body problem. In Section~\ref{sec:materials},  we discuss the role of materials and the CI $\mathbf{k}\cdot\mathbf{p}$ calculations on realistic devices. Section~\ref{sec:implications} is devoted to the analysis of the consequences of Wigner molecularization on two-qubit gate and readout performances, and Section~\ref{sec:conclusion} to our conclusions.

\begin{figure*}
\centering
  \includegraphics[width=\textwidth]{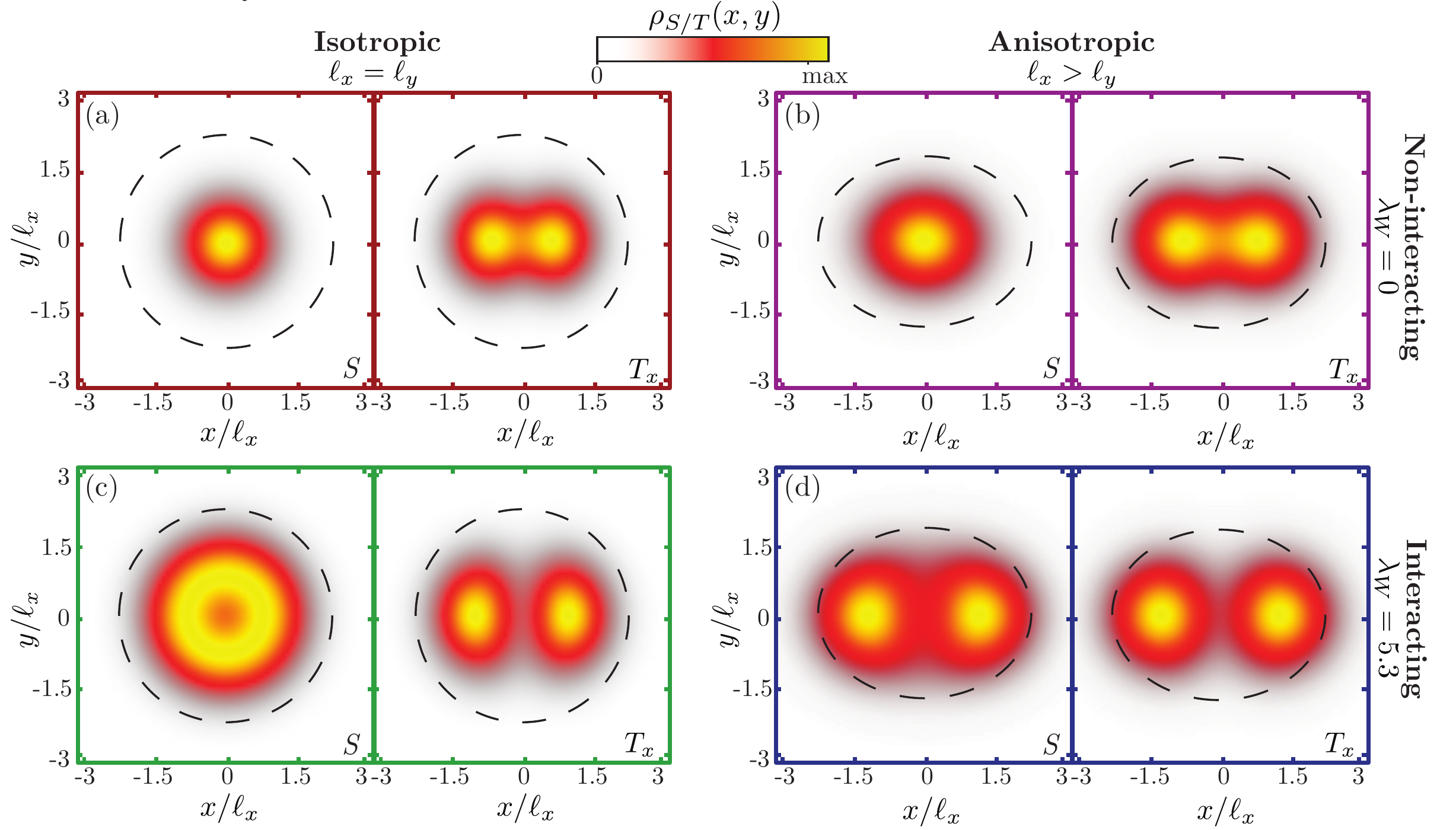}
\caption{Pairs of singlet $S$ and triplet $T_x$ ground-state densities for isotropic $\ell_x=\ell_y$ (left) and anisotropic $\ell_y=0.8\ell_x$ (right) dots, and for non-interacting ($\lambda_W=0$, top) and interacting ($\lambda_W=5.3$, bottom) particles. The black dashed ellipses with major axis $4.5\ell_x$ and minor axis $4.5\ell_y$ delineate the shape of the dots.}
\label{fig:wfns}
\end{figure*}

\section{Wigner molecularization in ``single-band'' harmonic potentials}\label{sec:model}

\subsection{Model}

We consider quantum dots strongly confined along the $z$ direction. We model confinement in the $(xy)$-plane by an anisotropic, harmonic 2D potential (Fig.~\ref{fig:schematics}). For simplicity, we make use of the single band approximation, which is expected to hold in many practical cases: ({\it i}) for holes, the ground state is generally of strong heavy-hole character, and is separated from the light-hole states by relatively large energy splittings~\citep{Maurand:2016p13575, VenitucciPRB2019, Hendrickx_Nature2020, hendrickx2021four, MichalPRB2021}; ({\it ii}) for electrons, the single band approximation is usually valid in materials (such as III-Vs) that have no valley degree of freedom. The impact of Coulomb interactions on the physics of low-lying valley states in Si or Ge has been addressed in Refs.~\citep{Jiang_Coulomb_PRB2013,ercan2021strong,ercan2021charge} (for 2D Si/SiGe dots) and will be briefly discussed at the end of Section~\ref{subsec:materials}. 

\correction{In the single band approximation, the motion of two particles in an isotropic harmonic potential is a problem of particular interest in physical chemistry known as Hooke’s atom~\citep{kestner1962study,kais1993density}. This problem has been largely investigated using both perturbative~\citep{white1970perturbation} and analytical methods for specific values of the confinement potential~\citep{TautPRA1993, zhu2005analytical}. In particular, Hooke's atom model is widely used to understand the formation of Wigner molecules under strong magnetic fields~\citep{yannouleas2002magnetic, szafran2004plane, szafran2004anisotropic}, their optical properties~\citep{peeters1990magneto}, and their vibrational modes~\citep{yannouleas2000collective}. In this work, we focus on the role of anisotropy in the formation of Wigner molecules and in the reduction of the singlet-triplet splitting. We analyze, in particular, the consequences of Wigner molecularization on the performances of spin qubits made of different materials relevant for semiconductor-based quantum technologies. The Hamiltonian of the anisotropic 2D Hooke's atom with material-dependent parameters is:} 
\begin{equation}
    H=\sum_{i=1}^2 \frac{\mathbf{p}^2_i}{2m}+\frac{1}{2}m\left(\omega_x^2x_i^2+\omega_y^2y_i^2\right)+\frac{e^2}{4\pi\varepsilon|\mathbf{r}_1-\mathbf{r}_2|}\,,
    \label{eq:H}
\end{equation}
where $\varepsilon=\varepsilon_r\varepsilon_0$ is the dielectric permittivity and $m$ is the in-plane effective mass. The confinement along $x$ and $y$ is characterized by the energies $\hbar\omega_x$ and $\hbar\omega_y$, or alternatively by the length scales $\ell_{x}=\sqrt{\hbar/(m\omega_{x})}$ and $\ell_{y}=\sqrt{\hbar/(m\omega_{y})}$. Given that the solutions for $\omega_x<\omega_y$ and $\omega_x>\omega_y$ are simply related by a permutation of the $x$ and $y$ axes, we assume $\omega_x\le\omega_y$ ($\ell_{x}\ge\ell_{y}$) without loss of generality. In the absence of Coulomb interactions, Eq.~\eqref{eq:H} is the Hamiltonian of a 2D harmonic oscillator. The singlet is then the doubly occupied $s$-like ground state, and the triplet is obtained by promoting one electron to the first excited state. Therefore, the singlet-triplet splitting is  $\Delta_{ST}^{(0)}=\hbar\omega_x$ (see Fig.~\ref{fig:schematics}a). In the presence of Coulomb interactions, the Hamiltonian is separable in the center of mass and relative coordinates $\mathbf{R}=(\mathbf{r}_1+\mathbf{r}_2)/2$ and $\mathbf{r}=\mathbf{r}_1-\mathbf{r}_2$. Namely, $H = H_{\mathbf R}+H_{\mathbf r}$, where:
\begin{subequations}
\begin{align}
   \label{eq:HR} H_{\mathbf R}& = \frac{\mathbf{P}^2}{4m}+m\left(\omega_x^2\,X^2+\omega_y^2\,Y^2\right)\,, \\  
   \label{eq:Hr} H_{\mathbf r}& = \frac{\mathbf{p}^2}{m}+\frac{m}{4}\left(\omega_x^2\,x^2+\omega_y^2\,y^2\right)+\frac{e^2}{4\pi\varepsilon|\mathbf r|}\,, 
\end{align}
\end{subequations}
with $\mathbf P$ and $\mathbf p$ the conjugate momenta to $\mathbf R$ and $\mathbf r$, respectively. The motion of the center of mass remains described by the Hamiltonian $H_{\mathbf R}$ of a non-interacting 2D harmonic oscillator, but with a doubled mass, which has no effect on the eigenenergies but shrinks the wave functions. The relative motion is also described by the Hamiltonian $H_{\mathbf r}$ of a 2D harmonic oscillator, with a halved mass and an extra Coulomb potential at the origin. By design, the even parity eigenstates of $H_{\mathbf r}$ are the singlets, while the odd parity eigenstates are the triplets (since they are, respectively, symmetric and anti-symmetric with respect to the exchange of particles $\mathbf{r}\to-\mathbf{r}$). In a simple perturbative picture, detailed in Appendix~\ref{app:perturbation}, one expects $\Delta_{ST}<\Delta_{ST}^{(0)}$ once the repulsive Coulomb interactions are turned on. Indeed, the singlet wave functions have a maximum at the position of the Coulomb singularity $\mathbf{r}=\mathbf{r}_1-\mathbf{r}_2=\mathbf{0}$, while the triplet wave functions have a node. Therefore, the ground-state singlet energy rises faster than the triplet energy when strengthening Coulomb interactions, which closes the gap $\Delta_{ST}$ (see Fig.~\ref{fig:schematics}b). 

The relevant energy scales of this problem are the orbital energy $E_\text{orb}=\hbar\omega_x$ and the Coulomb repulsion between two electrons separated by $\ell_x$, $E_\text{ee}=e^2/(4\pi\varepsilon\ell_x)$ \citep{ercan2021strong}. We  introduce the reduced coordinates $(x',\,y')=(x/\ell_x,\,y/\ell_x)$ to write down an unitless Hamiltonian $H'_{\mathbf r}=H_{\mathbf r}/E_\text{orb}$:
\begin{equation}\label{eq:dimensionlessH}
    H'_{\mathbf r}=-\partial^2_{x'}-\partial^2_{y'}+\frac{1}{4}\left(x'^2+\alpha ^2\,y'^2\right)+\frac{\lambda_W}{|\mathbf r'|}\,,
\end{equation}
where we have defined the Wigner ratio $\lambda_W=E_\text{ee}/E_\text{orb}$, which quantifies the interaction strength, and the anisotropy $\alpha=\omega_y/\omega_x=\ell_x^2/\ell_y^2$. The limit $\lambda_W\rightarrow 0$ corresponds to effectively turned-off Coulomb interactions either due to $\varepsilon\rightarrow\infty$ (complete screening of Coulomb interactions) or a small dot $\ell_x\rightarrow 0$ (where the kinetic energy prevails over Coulomb interactions), while large $\lambda_W$ is associated to a low $\varepsilon$ and/or a large dot. We solve Eq.~\eqref{eq:dimensionlessH} numerically to quantify the singlet-triplet splitting. We refer to Appendix~\ref{app:model} for details on the numerical solution, and to Appendix~\ref{app:model3d} for an extension to ``quasi-3D'' systems (2D electron/hole gas with finite thickness).

\begin{figure}
\centering
  \includegraphics[width=0.45\textwidth]{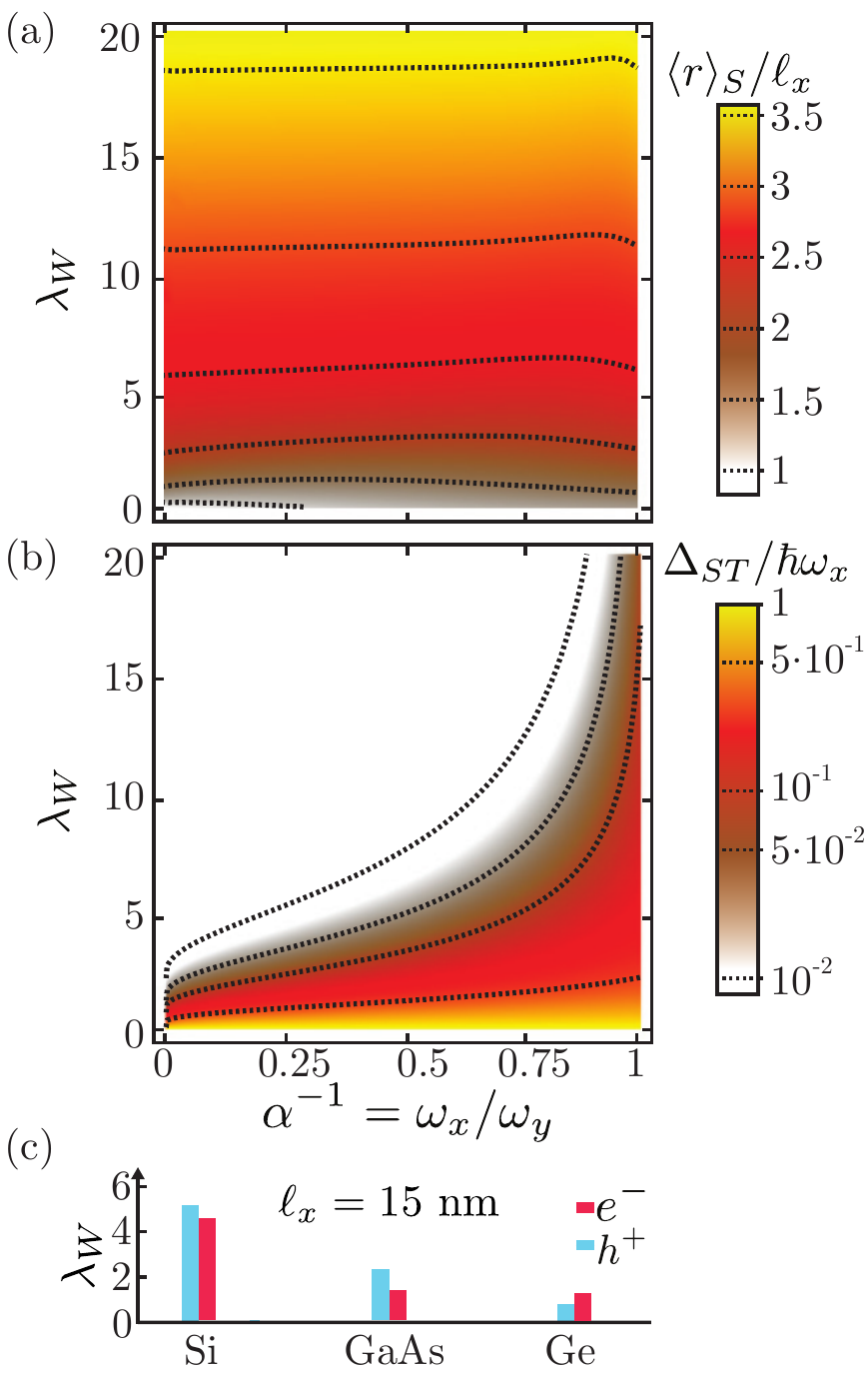}
\caption{Enlargement of the dot and suppression of the singlet-triplet splitting $\Delta_{ST}$ as a function of the inverse anisotropy\footnote{\correction{Although $\alpha=\omega_y/\omega_x=\ell_x^2/\ell_y^2$ is a natural measure of the anisotropy ($\alpha\in[1, \infty[$ consistently increasing with increasing $\ell_x>\ell_y$), Fig. \ref{fig:splittings-uless} is more readable when plotted as a function of $\alpha^{-1}$, which has a finite support $\alpha^{-1}\in]0, 1]$.}} $\alpha^{-1}=\omega_x/\omega_y$ and interaction strength $\lambda_W$. (a) Average distance $\langle r\rangle_S$ between the particles in the singlet state (in units of the major dot radius $\ell_x$). (b) Singlet-triplet splitting $\Delta_{ST}$ (in units of the single-particle orbital splitting $\hbar\omega_x$). Iso-level curves are shown as dashed lines. \correction{(c) Value of $\lambda_W=\ell_x/a_B$ for an electron (blue) or hole (red) dot with major radius $\ell_x=15$ nm in Si, Ge and GaAs (also see Table~\ref{tab:data}). These bars can be used to convert the $\lambda_W$ scale of panels (a) and (b) into a $\ell_x$ scale for each material.}} 
\label{fig:splittings-uless}
\end{figure}

\subsection{Anisotropy effects} 

To illustrate the strong impact that a finite anisotropy $\alpha>1$ has on the spatial properties of the ground-state singlet ($S$) and first excited triplet ($T$) states, we plot the corresponding two-particle densities in Fig.~\ref{fig:wfns}. These densities are defined as
$\rho_{S/T}(x_1,y_1)=2\int d^2\mathbf{r}_2\,| \psi_0(\mathbf{R})\varphi_{0/1}(\mathbf{r})|^2$, where $\psi_0(\mathbf R)$ is the Gaussian ground-state wave function of the center of mass Hamiltonian [Eq.~\eqref{eq:HR}], and $\varphi_{0/1}(\mathbf r)$ is the ground/first-excited wave function of Eq.~\eqref{eq:Hr}. Specifically, we compare the isotropic and anisotropic cases, with and without interactions.  

In the non-interacting and isotropic case ($\lambda_{W}=0$, $\alpha=1$), the singlet density is maximum at the center of the dot (doubly occupied $s$-like state, Figs.~\ref{fig:schematics}a and~\ref{fig:wfns}a), while the triplet density already has a dip (singly occupied $s$- and $p_x$-like states, degenerate with a singly occupied $p_y$ counterpart). Once the Coulomb repulsion is turned on, the dot swells to minimize the total energy. In the isotropic limit, the singlet density gets depleted at the center of the dot, owing to the competition between Coulomb interactions, which tend to keep the particles apart, and the harmonic confinement, which tends to bring the particles back to the origin (Figs.~\ref{fig:schematics}b and~\ref{fig:wfns}c). As a consequence, the two particles get engaged in a correlated ``dance'' around the center of the dot. The triplet $T_x$ shown in Fig.~\ref{fig:wfns}c is degenerate with a $T_y$ partner and remains qualitatively different from the singlet. 

This picture however breaks down in highly anisotropic dots, where Coulomb interactions dominate because the two particles cannot cross each other without coming exceedingly close. As a consequence, the carriers separate on both sides of the major axis of the dot, as shown in Fig.~\ref{fig:schematics}c and~\ref{fig:wfns}d. Accordingly, the spatial structure of the ground-state singlet approaches that of the first excited triplet state when increasing $\alpha$ and $\lambda_W$, in clear contrast to the isotropic case: the system shows the fingerprints of a double dot -- a ``Wigner molecule'' -- even though the harmonic potential landscape is actually that of a single dot. Since the separation of the particles can be understood as an hybridization between interacting $s$- and $p_x$-like states, it strongly destabilizes the singlet-triplet splitting, \correction{as discussed in the following paragraphs.}

\begin{figure}
\centering
  \includegraphics[width=0.4\textwidth]{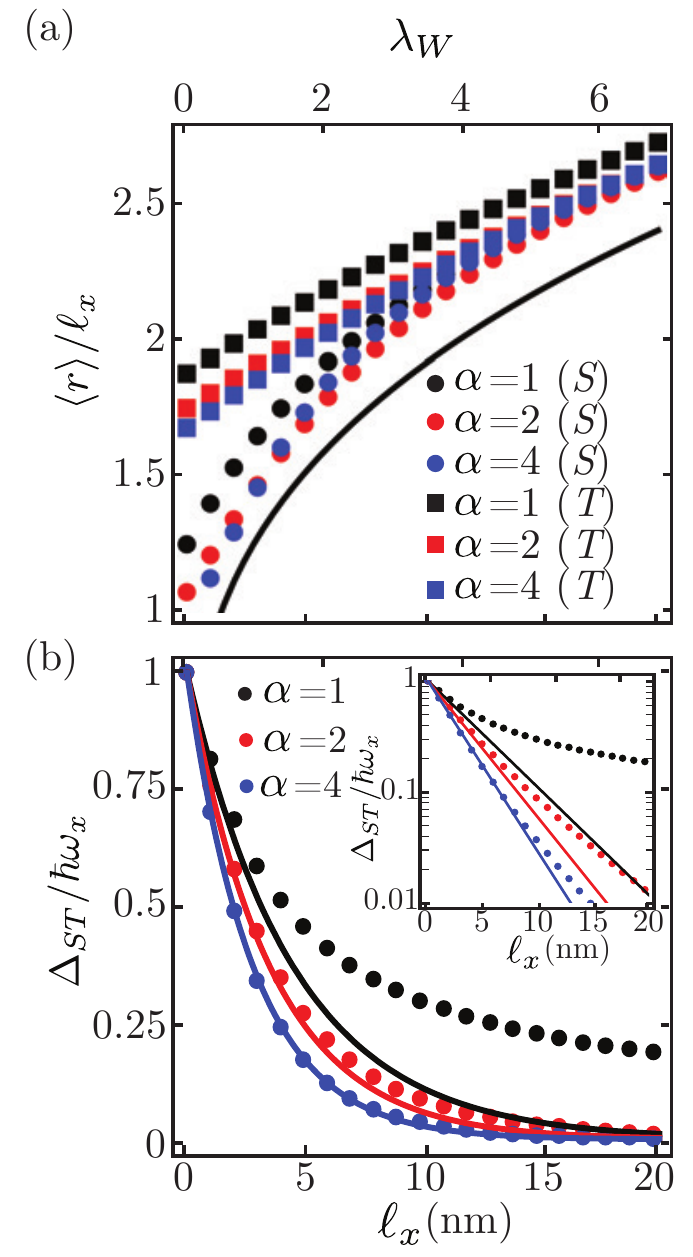}
\caption{Enlargement of the dot and singlet-triplet splitting for two holes in silicon. (a) Average inter-particle distance $\langle r\rangle$ in the singlet ($S$) and triplet ($T$) states as a function of $\ell_x$ for different dot anisotropies. The dots are the numerical calculations and the solid line is the classical estimate given by Eq.~\eqref{eq:interdistance}. (b) $\Delta_{ST}$ as a function of $\ell_x$ for the same dot anisotropies. The dots are the numerical calculations while the solid lines are the analytical approximations given by Eq.~\eqref{eq:DeltaSTfull}. In the inset, the same plot in logarithmic scale, showing the exponential trend for large $\alpha$. Note that these data can be translated to other materials using the $\lambda_W/\ell_x$ scale factors of Table~\ref{tab:data} and the Wigner ratio $\lambda_W$ on top.}
\label{fig:splittings}
\end{figure}

To get further insight into these trends, we plot the average distance $\langle r\rangle_{S/T}$ between the particles in the singlet and triplet states in Figures~\ref{fig:splittings-uless}a and \ref{fig:splittings}a. At small $\lambda_W$, $\langle r\rangle$ is slightly larger in the triplet than in the singlet state owing to Pauli exclusion principle (formation of an ``exchange hole'' in the pair correlation function of the triplet). When increasing $\lambda_W$, the dot swells and the strong Coulomb repulsion ultimately keeps the two particles equally separated in both the singlet and triplet states; $\langle r\rangle_S$ and $\langle r\rangle_T$ hence approach the same classical limit
\begin{equation}
    \langle r\rangle\approx\sqrt[3]{2\lambda_W}\ell_x\
    \label{eq:interdistance}
\end{equation}
that minimizes the Coulomb plus potential energy in Eq.~\eqref{eq:dimensionlessH} (the kinetic energy becoming irrelevant in the strongly interacting dot). As shown in Fig.~\ref{fig:splittings}a, the average inter-particle distance $\langle r\rangle$ is weakly dependent on the anisotropy $\alpha$ of the dot; yet the physics is qualitatively different for quasi-isotropic and highly anisotropic dots. In the first case, the particles are still free to move around the dot, although not independently from each other, while, in the second case, particles can only separate at both ends of the dot. 

Since the two particles come closer in the singlet than in the triplet state, the gap $\Delta_{ST}$ is expected to close even in the weakly interacting isotropic limit (Fig.~\ref{fig:schematics}b). However, the singlet-triplet splitting in a strongly anisotropic dot gets ultimately ruled by the residual exchange interactions between the two ``atoms'' of the Wigner molecule, and shall decay much faster when strengthening interactions. This trend is highlighted in Figs.~\ref{fig:splittings-uless}b and~\ref{fig:splittings}b. \correction{For a given confinement energy $\hbar\omega_x$}, the main general conclusion is that, while the singlet-triplet splitting is always suppressed by the interactions ($\Delta_{ST}\leq\hbar\omega_x$), this suppression is strongly enhanced by the anisotropy. In particular, $\Delta_{ST}\rightarrow 0$ in the pure 1D limit $\omega_y\gg \omega_x$. Additionally, we show in the inset of Fig.~\ref{fig:splittings}b that $\Delta_{ST}$ does, in fact, decrease exponentially with increasing $\lambda_W$, and the faster the larger the anisotropy $\alpha$, in agreement with the above considerations.

\subsection{Semi-analytical formula for $\Delta_{ST}$} 

In this section, we develop a first-order perturbation theory for the singlet-triplet splitting, and show how it can be extrapolated to describe the exponential suppression of $\Delta_{ST}$ with increasing $\lambda_W$, in the $\alpha\gg 1$ limit. We summarize the main steps here and refer to Appendix~\ref{app:perturbation} for details. Let $\phi_n(x')$ be the $n$-th eigenstate of the 1D harmonic oscillator. In the non-interacting limit, the singlet and triplet solutions of Eq.~\eqref{eq:dimensionlessH} are $\varphi_S^{(0)}(x',y')=\phi_0(x')\phi_0(y')$ and $\varphi_T^{(0)}(x',y')=\phi_1(x')\phi_0(y')$, respectively. The first-order correction to $\Delta_{ST}^{(0)}=\hbar\omega_x$ is then $\Delta_{ST}^{(1)}/\hbar\omega_x=\lambda_W[\bra{\varphi_T^{(0)}}1/r'\ket{\varphi_T^{(0)}}-\bra{\varphi_S^{(0)}}1/r'\ket{\varphi_S^{(0)}}]$, which evaluates to 
\begin{equation}
 \frac{\Delta_{ST}^{(1)}}{\hbar\omega_x}=-\sqrt{\frac{2}{\pi}}\frac{\sqrt{\alpha}}{\alpha-1}\left[\alpha \,\mbox K(1-\alpha)-\mbox E(1-\alpha)\right]\lambda_W\,,
 \label{eq:DeltaST1}
\end{equation}
where $\mbox K$ and $\mbox E$ are the elliptic integrals of the first and second kind, respectively. As illustrated in Fig.~\ref{fig:log} of Appendix~\ref{app:perturbation}, Eq.~\eqref{eq:DeltaST1}  diverges logarithmically when $\alpha\gg1$, which underlines the dominance of interactions in this limit. 

Noting on Fig.~\ref{fig:splittings}b that $\Delta_{ST}/\hbar\omega_x\approx\exp[-F(\alpha)\lambda_W]$, we can estimate the function $F(\alpha)\lambda_W\approx \Delta_{ST}^{(1)}/\hbar\omega_x$ from the power series expansion of the exponential. Conversely,
\begin{equation}
 \Delta_{ST}\approx\hbar\omega_x e^{-\Delta_{ST}^{(1)}/\hbar\omega_x}\,.
 \label{eq:DeltaSTfull}
\end{equation}
Even though this is not a rigorously derived formula, it shows a remarkable agreement with the numerical data shown in Fig.~\ref{fig:splittings}d. In particular, this expression is very well-behaved in the anisotropic limit $\alpha\gg 1$. It is, as expected, poorer in the isotropic limit, where the higher-order contributions from virtual excitations (dynamical correlations) become more relevant, but are not accounted for in the derivation of Eq.~\eqref{eq:DeltaST1}.

\section{Impact of materials and device layout}\label{sec:materials}

We now quantify the relevance of Wigner molecularization in usual semiconductor materials and in realistic quantum dot devices. 

\subsection{Materials}
\label{subsec:materials}

Semiconductor quantum dot qubits are, for the most part, hosted in Si, Ge and GaAs. In Eq.~\eqref{eq:H}, there are two material-dependent parameters: the in-plane effective mass $m$ and the relative dielectric permittivity $\varepsilon_r$. For given $\omega_x$ and $\omega_y$ (or, alternatively, $\ell_x$ and $\ell_y$), we expect, therefore, different singlet-triplet splitting renormalizations depending on the host material. In realistic devices, both $\alpha$ and $\lambda_W$ are tunable to some extent when the confinement is controlled by gate(s) since $\lambda_W$ is also a function of $\omega_x$. Specifically, the Wigner ratio is proportional to the harmonic size of the dot along the weakest confinement axis: $\lambda_W=\ell_x/a_B$, where $a_B=4\pi\hbar^2\varepsilon/(me^2)$ is the effective Bohr radius. We give the scaling factor $\lambda_W/\ell_x=1/a_B$ for both electrons and holes in the three considered materials in Table~\ref{tab:data}. This scaling factor and the anisotropy $\alpha$ are the only inputs needed to establish where a specific quantum dot in an experimental setup is situated in the parameter space of Fig.~\ref{fig:splittings-uless}. 

\begin{figure*}
\centering
  \includegraphics[width=0.8\textwidth]{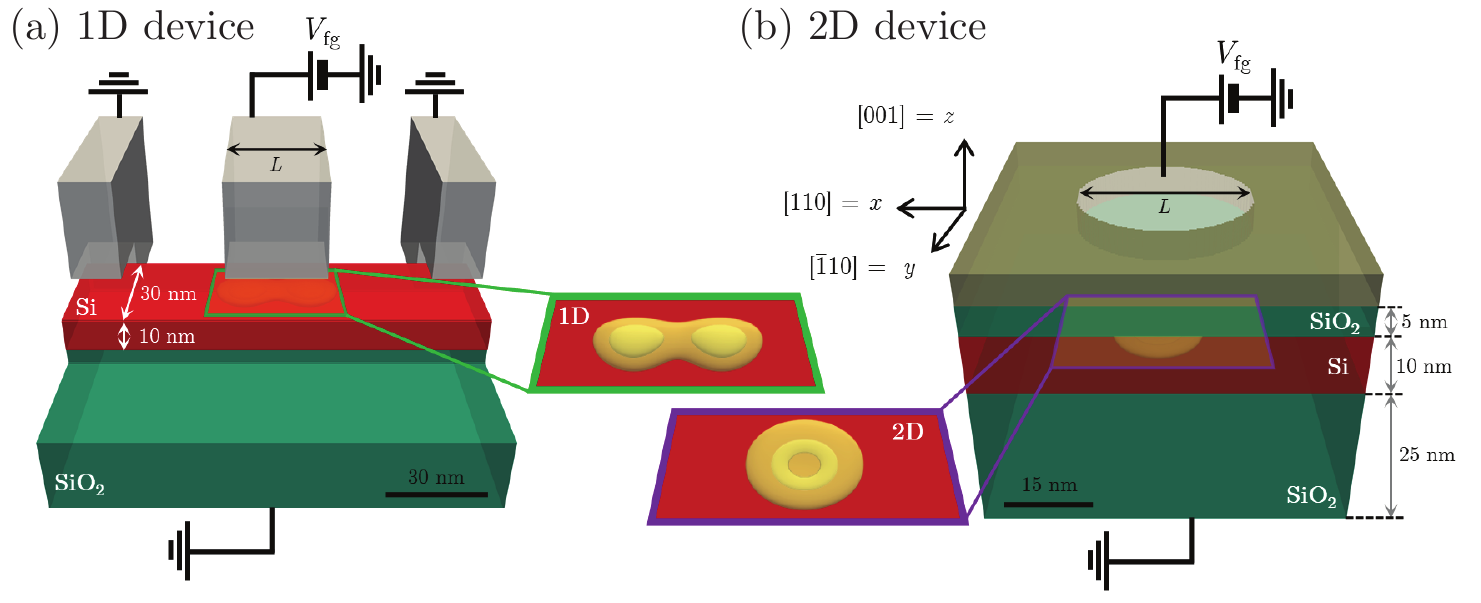}
\caption{Illustration of the realistic device simulations. (a) Quasi-1D quantum dot in a silicon nanowire channel. Silicon is colored in red, SiO$_2$ in green and the gates in gray. The channel is 30 nm wide and 10 nm thick. The dot is confined in the corner of the channel by the central gate with length $L$. The 5 nm thick SiO$_2$ around the wire and the Si$_3$N$_4$ spacers between the gates have been removed from the image to disclose the silicon channel. The calculated two-hole singlet density is plotted under the gate and enlarged in the inset for convenience ($V_{\rm fg}=-55$ mV). (b) 2D quantum dot. The dot is now confined by the isotropic potential of a circular gate. The calculated two-hole singlet density is also plotted in the device and inset ($V_{\rm fg}=-55$ mV). Note that the singlet densities in both devices are qualitatively similar to those of the simple harmonic model (Fig.~\ref{fig:wfns}).}
\label{fig:devices}
\end{figure*}

\correction{We illustrate the relevant dot sizes in Fig.~\ref{fig:splittings-uless}c as scale bars for electrons and holes in the three materials.} They outline the strength of the interactions for a reference dot radius $\ell_x=15$ nm. The length of these bars and the values of $\lambda_W/\ell_x$ listed in Table~\ref{tab:data} support the same conclusion: Si dots are way more impacted by Coulomb interactions than GaAs dots, which are themselves more affected than Ge dots, especially in the case of holes. These trends essentially reflect the differences between the heavy-hole masses of these materials: the heavier the carriers, the stronger the Coulomb correlations. As illustrated in Fig.~\ref{fig:splittings}, silicon dots with realistic sizes $\ell_x>5$ nm swell significantly when doubly occupied by holes ($\langle r\rangle/\ell_x\gg 1$), while the singlet-triplet gap is much reduced ($\Delta_{ST}/\hbar\omega_x<0.2$), unless the dot is almost perfectly isotropic. In comparison, Ge dots are more resilient to interactions, as they can be made up to six times larger than Si dots before undergoing a comparable reduction of $\Delta_{ST}/\hbar\omega_x$. GaAs dots lie in between, and can be made twice larger than Si dots for a given $\Delta_{ST}/\hbar\omega_x$. The trends are softer for electrons, but GaAs and Ge dots can still be three times larger than Si dots for a similar decrease of $\Delta_{ST}/\hbar\omega_x$ (see however discussion below on the valley degrees of freedom). A larger $\lambda_W$ also implies a wider stretching of the dot by electron-electron interactions, as shown by Eq.~\eqref{eq:interdistance} and Fig.~\ref{fig:splittings}a, although the differences between materials are smaller since $\langle r\rangle$ is only $\propto \lambda_W^{1/3}$. Larger dots may have stronger dipole moments, which can be leveraged for faster electric dipole spin resonance, but can also make the qubits more sensitive to disorder and noise \citep{Biel:unpublished}. Anyhow, the effects of interactions are weaker in small and isotropic dots, which implies that dots confined in nanowires are particularly susceptible. 

To conclude, we would like to comment on Wigner molecularization in the presence of the valley degree of freedom for electrons in Si and Ge. When the triplet state is based on a valley excitation with the same envelope as the ground-state but a different Bloch function, the singlet-triplet gap $\Delta_{ST}$ is rather robust to Coulomb interactions \citep{ercan2021charge}, because mixing the two valleys does not separate the electrons on either side of the dot \citep{Jiang_Coulomb_PRB2013}. In that case the present theory holds for the first triplet of orbital nature, which may ultimately become the lowest-lying triplet in anisotropic or large enough dots. The valley singlet-triplet gap is nonetheless more sensitive to Coulomb interactions in the presence of strong disorder-induced valley-orbit coupling, which mixes the valley and orbital degrees of freedom. This problem is discussed in particular in Refs.~\citep{ercan2021strong,ercan2021charge} \correction{and in Appendix \ref{app:wignervalleys}}. Singlet/triplet/quintet/septet gaps much smaller than the expected single-particle valley and orbital splittings have for example been measured in different anisotropic silicon-on-insulator (SOI) dots structurally similar to those discussed in the next section \cite{lundberg2020spin,noteFGZ}. Although some of these gaps were measured in the many-electron regime, Coulomb correlations likely contribute to the overall compression of the excitation spectrum.

\begin{table}
\begin{centering}
\begin{tabular}{||c|c|c|c||c|c|}
\hline
  & ~Si~ & ~Ge~ & ~GaAs~ \tabularnewline
\hline
$\gamma_1$ & ~4.29~  & ~13.38~  & ~6.98~ \tabularnewline
\hline
$\gamma_2$ & ~0.34~  & ~4.24~  & ~2.06~ \tabularnewline
\hline
$m^{(h)}/m_0$ & ~0.21~  & ~0.06~  & ~0.11~ \tabularnewline
\hline
$m^{(e)}/m_0$ & ~0.19~ & ~0.08$^*$~ & ~0.066~ \tabularnewline 
\hline
$\varepsilon_r$ & ~11.7~ & ~16.2~ &  ~12.9~ \tabularnewline
\hline
~$\lambda_W^{(h)}/\ell_x$ (nm$^{-1}$)~ & ~0.35~ & ~0.06~ & ~0.16~  \tabularnewline
\hline
~$\lambda_W^{(e)}/\ell_x$ (nm$^{-1}$)~ & ~0.31~ & ~0.09~ & ~0.10~  \tabularnewline
\hline
\end{tabular}
\par\end{centering}
\caption{\label{tab:data} Physical parameters of the three considered materials and scaling factor $\lambda_W/\ell_x=1/a_B$ between the major dot radius $\ell_x$ and the Wigner ratio $\lambda_W$. $m^{(e)}$ is the in-plane electron mass, $m^{(h)}=m_0/(\gamma_1+\gamma_2)$ is the in-plane heavy-hole mass (with $\gamma_1$ and $\gamma_2$ the Luttinger parameters), and $\varepsilon_r$ is the static dielectric constant. All dots are assumed to be strongly confined along $z=[001]$ except ($^*$) for electrons in Germanium ($z=[111]$).}
\end{table}

\subsection{Simulations in realistic devices}
\label{subsec:devices}

\begin{figure}
\centering
  \includegraphics[width=0.4\textwidth]{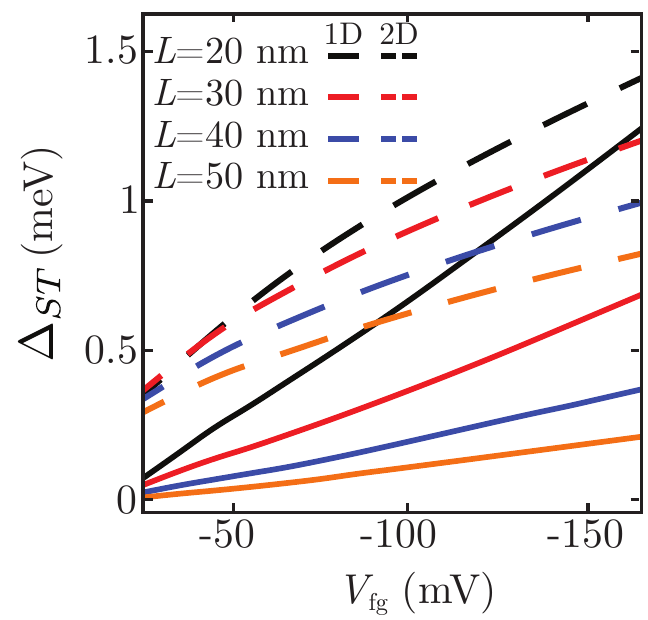}
\caption{Singlet-triplet splitting $\Delta_{ST}$ as a function of the front gate voltage $V_{\rm fg}$ for different gate lengths $L$ in realistic 1D and 2D devices (two-hole quantum dots in silicon).}
\label{fig:splittings-meV}
\end{figure}

\begin{figure*}
\centering
  \includegraphics[width=\textwidth]{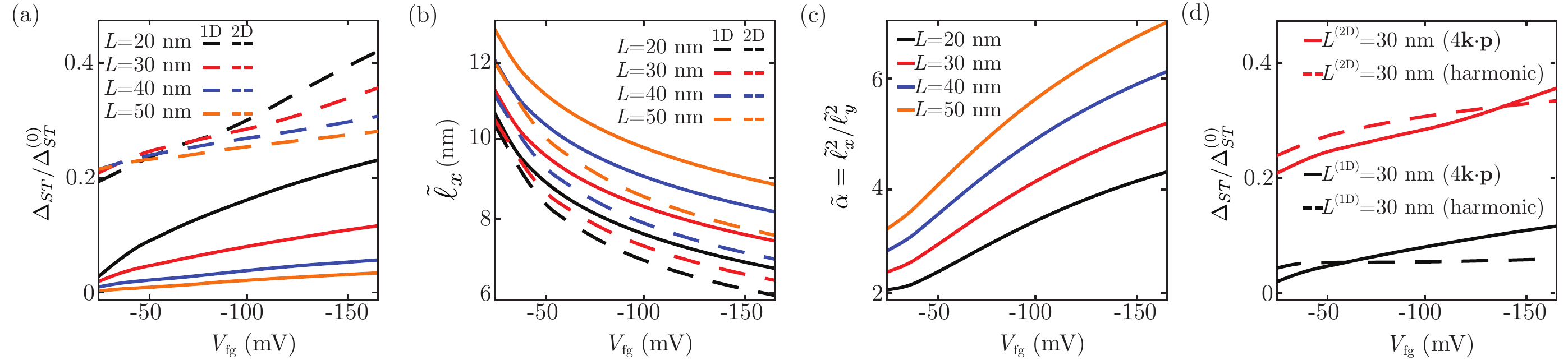}
\caption{Full configuration interaction results in realistic two-hole quantum dots in silicon. (a) Normalized singlet-triplet splitting $\Delta_{ST}/\Delta_{ST}^{(0)}$ as a function of the front gate voltage $V_{\rm fg}$ for different gate lengths $L$ in the 1D and 2D devices of Fig.~\ref{fig:devices}. $\Delta_{ST}^{(0)}$ is the orbital splitting in the single-hole configuration. (b) Effective major dot radius $\tilde{\ell_x}$ in the single-hole ground state of the 1D and 2D devices. (c) Effective dot anisotropy $\tilde{\alpha}=\tilde{\ell}_x^2/\tilde{\ell}_y^2$ in the single-hole ground state of the 1D devices. (d) Comparison between the full CI splitting $\Delta_{ST}/\Delta_{ST}^{(0)}$ and the harmonic 2D model with the effective length $\tilde{\ell}_x$ and anisotropy $\tilde{\alpha}$ from panels (b) and (d) ($L=30$ nm in both 1D and 2D devices).}
\label{fig:splittings-devices}
\end{figure*}

Now that we have established how the singlet/triplet states and energies are renormalized by Coulomb interactions in an ideal harmonic potential, we investigate how Wigner molecularization affects more realistic devices relevant for experiments. \correction{We consider anisotropic quasi-1D quantum dots in a silicon nanowire channel, and isotropic 2D quantum dots in a silicon thin film (Fig.~\ref{fig:devices}). The film or nanowire is embedded in SiO$_2$, as in SOI devices~\cite{de2016soi}. We focus on holes, that have no valley degree of freedom, and illustrate the behavior of valley triplets in Appendix \ref{app:wignervalleys}.}

The 1D dots are confined in the corner of a $[110]$-oriented, rectangular silicon wire by a gate with length $L=20-50$ nm overlapping half of the channel (Fig.~\ref{fig:devices}a) ~\cite{VoisinNanoLett2014, BetzNL2015}. The width of the wire is $W=30$ nm [$(001)$ facets] and its height is $H=10$ nm [$(1\bar{1}0)$ facets]. The wire is lying on a 25 nm thick buried oxide and is insulated from the gate by 5 nm of SiO$_2$ ($\varepsilon_r=3.9$). The substrate below the buried oxide is grounded. Additional gates 30 nm on the left and right mimic neighboring dots and are also grounded in this study. The whole device is embedded in Si$_3$N$_4$ ($\varepsilon_r=7.5$). The 2D dots are confined in a $(001)$-oriented, 10 nm thick silicon film by a circular gate with diameter $L=20-50$ nm and the same stack of oxides as in the 1D dots (Fig.~\ref{fig:devices}b). The size of the dots can be further controlled by the voltage $V_{\rm fg}$ applied to the central gate. The more negative is $V_{\rm fg}$, the smaller are the 1D and 2D dots ($\ell_x$ and $\lambda_W$ decrease), and the stronger is the anisotropy in 1D dots ($\ell_x/\ell_y$ increases). The ranges of bias voltages and gate lengths explored in this work are typical for SOI devices and for silicon quantum dots~\cite{de2016soi}.

The potential landscape in the devices is computed with a finite volumes Poisson solver and the single-hole wave functions with a finite differences 4 bands $\mathbf{k}\cdot\mathbf{p}$ model (with the Luttinger parameters of Table~\ref{tab:data} and $\gamma_3=1.45$ in Si). The calculated single-hole states are then used as input to a full CI solver for the two hole wave functions (see Appendix~\ref{app:fullci} for details). The Coulomb integrals are calculated with the same Poisson solver as the potential landscape and are thus screened by the dielectrics and the gates. The silicon oxide tends, therefore, to enhance Coulomb repulsion ($\varepsilon_r$ lower than in pure Si), while the metal gate tends, on the opposite, to reduce both the strength and the range of the interactions.

The results for the 1D and 2D devices are shown in Figs.~\ref{fig:splittings-meV} and~\ref{fig:splittings-devices}. The absolute singlet-triplet
splitting $\Delta_{ST}$ is plotted as a function of $V_{\rm fg}$ in Fig.~\ref{fig:splittings-meV}. It remains greater than $\approx 0.5$\,meV in the 2D devices, but can be lower than $100\,\mu$eV in the 1D devices, where it also shows a stronger dependence on gate voltage and gate length. The renormalization of the singlet-triplet gap by Coulomb interactions can be appreciated in Fig.~\ref{fig:splittings-devices}a, where $\Delta_{ST}$ is normalized with respect to the orbital splitting $\Delta_{ST}^{(0)}$ in the non-interacting dot. For the gate lengths and biases considered here, $\Delta_{ST}/\Delta_{ST}^{(0)}$ ranges from $\approx 0.2$ to $\approx 0.4$ in 2D devices, but hardly reaches 0.2 in 1D devices. These trends are qualitatively consistent with the discussion of Section~\ref{sec:model}: the bare singlet-triplet splitting $\Delta_{ST}^{(0)}$ is heavily suppressed by interactions, especially in anisotropic 1D devices. As expected, $\Delta_{ST}/\Delta_{ST}^{(0)}$ decreases with increasing gate length, and with increasing $V_{\rm fg}\to 0$ (namely, with softening confinement). As the singlet-triplet splitting is electrically tunable, it can be controlled to some extent but is, on the other hand, sensitive to charge fluctuations. Altogether, the calculated suppression of $\Delta_{ST}/\Delta_{ST}^{(0)}$ can be considered strong, which highlights the necessity to account properly for Coulomb interactions in the description of realistic quantum dot systems. 

To provide a quantitative analysis of the geometrical deformation of the dot caused by the interactions, we define the ``effective'' dot sizes $\tilde{\ell}_x=\sqrt{2(\langle x^2\rangle-\langle x\rangle^2)}$ and $\tilde{\ell}_y=\sqrt{2(\langle y^2\rangle-\langle y\rangle^2)}$, where the expectations values are computed in the single hole ground state. These effective dot sizes coincide with $\ell_{x}=\sqrt{\hbar/(m\omega_{x})}$ and $\ell_{y}=\sqrt{\hbar/(m\omega_{y})}$ when evaluated in the ground state of a harmonic confinement potential. The effective $\tilde{\ell}_x$ is plotted in Fig.~\ref{fig:splittings-devices}b, along with the effective dot anisotropy $\tilde{\alpha}=\tilde{\ell}_x^2/\tilde{\ell}_y^2$ in the 1D devices (Fig.~\ref{fig:splittings-devices}c). As expected, $\tilde{\ell}_x$ decreases with decreasing gate size and increasing electric confinement (increasingly negative $V_{\rm fg}$). The anisotropy of the 1D dots is significant and is enhanced by the electric confinement in the corners ($\tilde{\alpha}>6$ at large $V_{\rm fg}<0$). The growing anisotropy is, however, outweighted by the contraction of the dot with more negative $V_{\rm fg}$ (smaller effective $\tilde{\ell}_x$ and $\tilde{\lambda}_W\simeq 0.35\tilde{\ell}_x$), which leads to a net increase of $\Delta_{ST}/\Delta_{ST}^{(0)}$. Finally, the singlet-triplet splitting $\Delta_{ST}/\Delta_{ST}^{(0)}$ obtained with the harmonic confinement model using $\tilde{\ell}_x$ and $\tilde{\ell}_y$ as input is compared to the full CI result in Fig.~\ref{fig:splittings-devices}d, for a specific gate length and diameter. The agreement is excellent for the 2D dots despite the different Coulomb kernels (the interaction in realistic devices being unscreened by the dielectrics, but screened by the gate). The agreement is somewhat poorer in the 1D case, though still quantitatively reasonable. The discrepancies result in part from the strong confinement along $y$, which is neither harmonic nor softer than the confinement along $z$ (as assumed in the model), and which significantly mixes heavy- and light-hole components (not accounted for in the single band approximation).

\correction{Similar conclusions can be reached for the orbital triplet in electron devices (see Appendix \ref{app:wignervalleys}). As discussed in section \ref{subsec:materials}, the valley splitting is robust to Coulomb interactions but the orbital singlet-triplet gap is strongly reduced in 1D silicon devices, and can even be smaller than the valley splitting in anisotropic or large enough dots.} These results confirm that the renormalization of the singlet-triplet splitting can be very large in realistic silicon devices, especially if the dots are anisotropic by design. We now discuss the implications of these results for the performances of the qubits.

\section{Discussion: Implications for spin qubits}\label{sec:implications}

The presence of low-lying triplet states affects the performances of quantum dot devices such as spin-qubits, in particular when doubly occupied single dot configurations are relevant. This happens, in particular, in two situations: when performing Pauli spin blockade (PSB) readout and when leveraging exchange interactions.

\begin{figure*}
\centering
  \includegraphics[width=.9\textwidth]{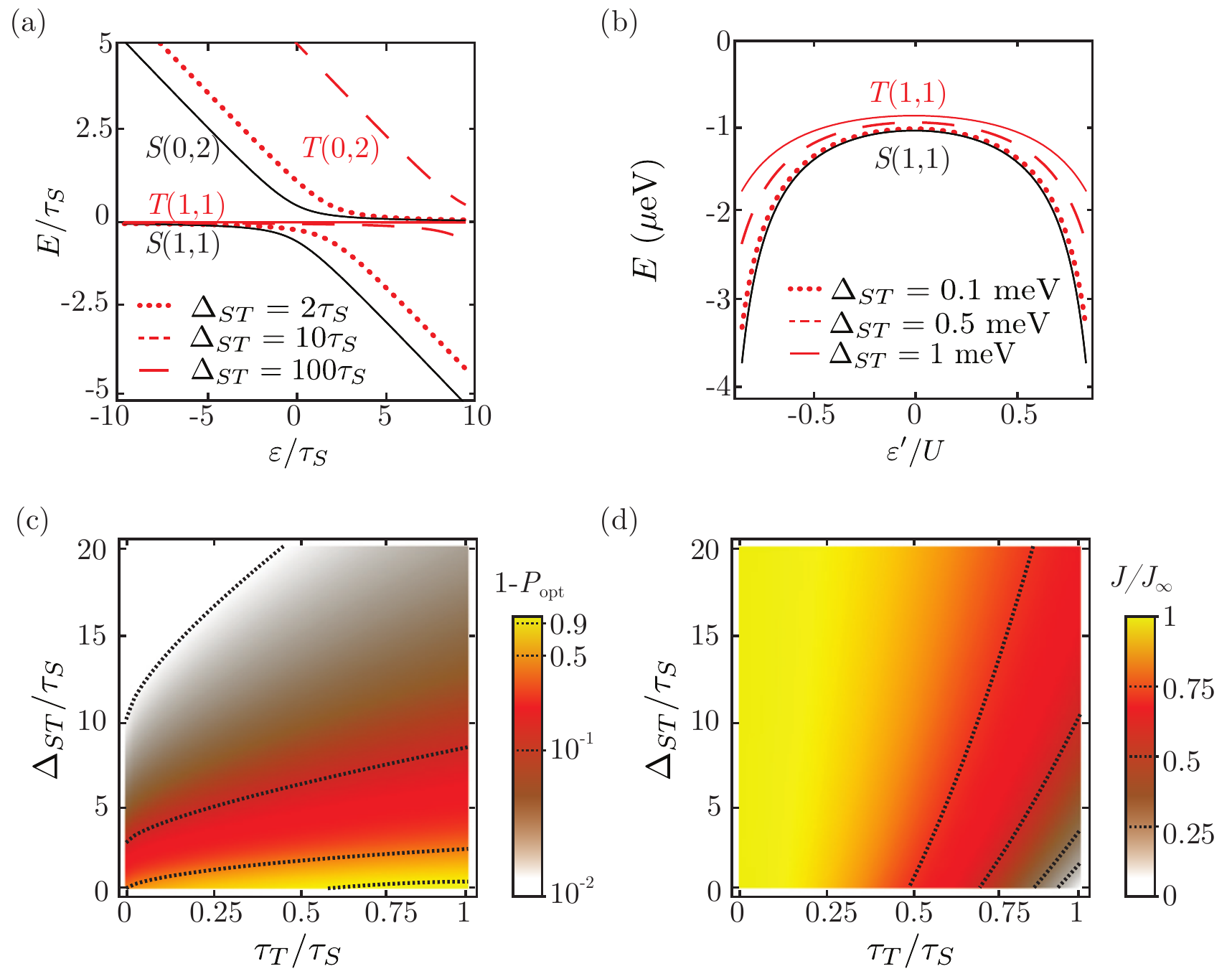}
\caption{Effects of the reduction of the singlet-triplet gap on the PSB readout and exchange interactions. \correction{(a) Energy diagram of the spin-zero states near the singlet $(1,1)/(0,2)$ anti-crossing at $\varepsilon=0$ with all energies normalized with respect to $\tau_S$. The black lines label the singlet states, while the red lines label the triplet states. For all cases we set $\tau_S=\tau_T$. The case $\Delta_{ST}=100\tau_S$ is plotted with solid lines, the case $\Delta_{ST}=10\tau_S$ with dashed lines, and the case $\Delta_{ST}=2\tau_S$ with dotted lines. Note the finite bending of the dotted and dashed $T(1,1)$ states as compared to the solid $T(1,1)$ state. (b) Energy diagram of the lowest singlet and triplet $(1,1)$ states centered at the SOP $\varepsilon'=0$ with $\tau_S=\tau_T=50$ $\mu$eV and $U=5$ meV. The black lines label the singlet state, while the red lines label the triplet states for different $\Delta_{ST}$. Note that a triplet state decoupled from the (0,2)-(2,0) configurations ($\Delta_{ST}\rightarrow\infty$ or $\tau_T=0$) would have zero energy.} (c) Probability of readout error $1-P_{opt}$ at the optimal readout point as a function of the triplet tunnel coupling $\tau_T$ and the singlet-triplet gap $\Delta_{ST}$, both given in units of the singlet tunnel coupling $\tau_S$. (d) Exchange coupling $J$ normalized with respect to the exchange interaction $J_\infty$ ($\Delta_{ST}\rightarrow\infty$), as a function of $\tau_T/\tau_S$ and $\Delta_{\rm ST}/\tau_S$. The detuning is $\varepsilon=-10\tau_S$.}
\label{fig:consequences}
\end{figure*}

\subsection{Pauli spin blockade readout}

Pauli spin blockade in double dots is a widely used mechanism for the readout of spin qubits. One of the two dots hosts the qubit, while the other one (the readout dot) is held in a fixed spin state. To measure the qubit, the double dot is detuned so as to enable spin dependent tunneling between the qubit and readout dots. If the double dot is in the singlet $S(1,1)$ state, tunneling is allowed and the transition to (or electric dipole with) the singlet $S(0,2)$ state can be detected by a charge sensing technique \citep{Petta:2004p1586, Johnson:2005p483}. However, if the double dot is in a triplet $T(1,1)$ state, tunneling to the doubly occupied $S(0,2)$ ground state is forbidden by Pauli exclusion principle.

The performance of this readout mechanism is limited by the presence of low-lying $T(0,2)$ states that allow for undesirable $T(1,1)/T(0,2)$ mixing. Indeed, the triplet $T(0,2)$ can also couple to the $(1,1)$ states if they are too close to the singlet $S(0,2)$; then a spurious transition from $T(1,1)$ to $T(0,2)$ during readout will be confused with the monitored $S(1,1)\to S(0,2)$ transition and spoil the measurement. PSB is therefore expected to be robust in small isotropic dots, where $\Delta_{ST}\approx\hbar\omega_x\gg\tau$ ($\tau$ being the typical tunnel coupling), and to become unreliable in large or anisotropic enough dots where $\tau\approx\Delta_{ST}$. We characterize the readout error here and discuss how to mitigate it.

To quantify the readout error we make the following assumptions: tunneling between singlet states, and between triplet states with the same spin projection is allowed, but tunneling between singlet and triplet states, and between triplet states with different spin projections is forbidden. Singlet-triplet tunneling is actually possible in the presence of spin-orbit coupling, but is in principle much weaker than the spin-conserving processes and is of minor importance for the question addressed here. We also assume that the Zeeman splitting remains much smaller than $\Delta_{ST}$ and thus neglect the action of the magnetic field on the triplet states. Within this approximation, the effective Hamiltonian of the $\{S(1,1), S(0,2), T(1,1), T(0,2)\}$ states reads:
\begin{equation}
H=\begin{pmatrix}
0 & \tau_S & 0 & 0 \\
\tau_S & -\varepsilon & 0 & 0 \\
0 & 0 & 0 & \tau_T \\
0 & 0 & \tau_T & -\varepsilon + \Delta_{ST}
\end{pmatrix}\,,
\label{eq:psbH}
\end{equation}
where $\varepsilon$ is the $S(1,1)/S(0,2)$ detuning, $\Delta_{ST}$ the singlet-triplet splitting in the $(0,2)$ configuration, $\tau_S$ the singlet-singlet tunneling, and $\tau_T$ the triplet-triplet tunneling. The resulting energy diagram is plotted in Fig.~\ref{fig:consequences}a, \correction{ for different $\Delta_{ST}/\tau_S$ ratios representative of various scenarios compatible with Fig.~\ref{fig:splittings-meV} assuming $\tau_S=\tau_T$ in the $\gtrsim 10$ $\mu$eV range}. When performing PSB readout, $\varepsilon$ is swept from the $(1,1)$ to the $(0,2)$ charge configuration; optimizing readout amounts to maximize the probability $P_S$ of a $S(1,1)\to S(0,2)$ transition while minimizing the probability $P_T$ of a $T(1,1)\to T(0,2)$ transition. We assume for simplicity that $\varepsilon$ can be swept adiabatically, so that $P_ {S}$ and $P_T$ are respectively the admixtures of $S(0,2)$ into $S(1,1)$ and $T(0,2)$ into $T(1,1)$ at the readout point. We get from the eigenvectors of Eq.~\eqref{eq:psbH}:
\begin{subequations}
\begin{align}
    P_S&=\frac{1}{2}+\frac{\varepsilon}{2\sqrt{\varepsilon^2+4\tau_S}}\,, \\
    P_T&=\frac{1}{2}+\frac{\varepsilon-\Delta_{ST}}{2\sqrt{(\varepsilon-\Delta_{ST})^2+4\tau_T}}\,.
    \label{eq:stprobs}
\end{align}
\end{subequations}
We hence need to maximize the quantity $P=P_S-P_T$. An optimal readout point $\varepsilon_\text{opt}$, and an optimal $P_\text{opt}=P_S(\varepsilon_\text{opt})-P_T(\varepsilon_\text{opt})$ can be found for any $\tau_S$, $\tau_T$, and $\Delta_{ST}$. The optimal readout point typically lies in the $0\le\varepsilon_\text{opt}\le\Delta_{ST}$ range. 

In Fig.~\ref{fig:consequences}c, we plot the readout error $1-P_{\rm opt}$ as a function of $\tau_T$ and $\Delta_{ST}$, both given in units of $\tau_S$. We restrict ourselves to the range $\tau_T\lesssim\tau_S$, as suggested by test calculations on realistic devices discussed in Appendix~\ref{app:tcoupling}. As expected, the readout error decreases when increasing $\Delta_{ST}$ and decreasing $\tau_T$, since both suppress the $T(1,1)/T(0,2)$ mixing. $\Delta_{ST}$ must, therefore, be made much larger than $\tau_T$ to achieve high-fidelity readout. 

In practice, $\tau_S$ and $\tau_T$ can be tuned electrically by a tunnel gate that controls the inter-dot space. However, the ratio $\tau_T/\tau_S$ is likely difficult to adjust, as the tunnel gate will in general act in a similar way on $\tau_S$ and $\tau_T$. The ratio $\Delta_{ST}/\tau_T$ is {\it a priori} much easier to control: either the quantum dot gate is biased to reshape the confinement and increase $\Delta_{ST}$, or the tunnel gate tuned to decrease $\tau_T$. However, reducing $\tau_S$ and $\tau_T$ too much will ultimately break adiabaticity along the $(1,1)\to(2,0)$ transition, thereby compromising the visibility of the singlet state.

As an example, assuming $\tau_T=\tau_S=\tau$, achieving $1-P_{\rm opt}=10^{-2}$ calls for $\Delta_{ST}>28\tau$. With typical readout tunnel couplings $\tau\gtrsim 10$ $\mu$eV \cite{Simmons:2009p3234,PRXQuantum.2.010303,ezzouch2021dispersively}, $\Delta_{ST}$ must, therefore, be at least $\simeq 300$ $\mu$eV, a value that may not be so easy to achieve in an anisotropic quantum dot. 

\correction{Finally, we would like to emphasize that the ratio $\tau_T/\tau_S$ can be strongly modulated by valley blockade effects in the conduction band of silicon \cite{Hao:2014p3860}. In the ideal situation where the two dots share the same valley states, and valley blockade is complete, $\tau_T\simeq 0$ for the valley triplet, so that the relevant $\Delta_{ST}$ and $\tau_T$ are those of the orbital triplet.}

\subsection{Exchange coupling}

Exchange coupling is the main mechanism used to drive two-qubit gates between spin qubits in the $(1,1)$ charge configuration~\citep{Loss:1998p120}. Another important consequence of the presence of a low-lying triplet state is the renormalization of this interaction. In essence, the tunnel coupling between neighboring dots bends down the singlet energy when detuning closer to the $S(1,1)/S(0,2)$ anti-crossing, which splits the singlet $S(1,1)$ from the triplet $T(1,1)$ states by an exchange energy $J$ (see Fig.~\ref{fig:consequences}a). This exchange energy can be controlled by adjusting the tunnel coupling and/or the detuning between the dots. A low-lying triplet $T(0,2)$ state can, however, also couple to the $T(1,1)$ states. The triplet $T(1,1)$ states then bend down similarly to the singlet state, so that the net value of $J$ is renormalized. We now quantify this effect as a function of the tunnelings $\tau_{S/T}$ and the singlet-triplet gap $\Delta_{ST}$ in the $(0,2)$ configuration, \correction{first at small detuning, then near the so-called symmetric operating point.}

\subsubsection{\correction{Small detuning}.}
We first focus our analysis on the ``small'' detuning range $-U\ll\varepsilon<0$ where the $(1,1)$ states only admix the $(0,2)$ states ($U$ being the intra-dot charging energy). The exchange coupling $J$ is defined as the difference between the ground triplet and singlet state energies. In the limit $\Delta_{ST}\rightarrow\infty$, we recover from the eigenvalues of Eq.~\eqref{eq:psbH} the usual expression of $J$ near the $(1,1)/(0,2)$ anti-crossing:
\begin{equation}
    J_\infty = \frac{1}{2}\left(\varepsilon+\sqrt{\varepsilon^2+4\tau_S^2}\right)\,.
    \label{eq:Jinfty}
\end{equation}
We have, however, assumed no bending of the triplet energy so far. Once we account for a finite $\Delta_{ST}$, the exchange coupling becomes:
\begin{equation}
    J=J_\infty-\frac{1}{2}\left(\varepsilon-\Delta_{ST}+\sqrt{(\varepsilon-\Delta_{ST})^2+4\tau_T^2}\right)\,.
    \label{eq:J02}
\end{equation}
The net exchange interaction is therefore reduced by the tunneling between triplet states. Such a decrease of $J$ slows down exchange-based two-qubit gates. The correction $\Delta J=J_\infty-J$ tends to zero when $\Delta_{ST}\rightarrow\infty$, and, more interestingly, when $\tau_T\rightarrow 0$ (see discussion below). 

To reduce the fluctuations of $J$ due to charge noise, it is desirable to operate away from the $(1,1)/(0,2)$ anti-crossing ($\varepsilon/\tau_S\ll 0$). Then, 
\begin{equation}
\frac{J}{J_\infty}\simeq 1-\left(\frac{\tau_T}{\tau_S}\right)^2\frac{\varepsilon}{\varepsilon-\Delta_{ST}}\,.
\end{equation}
As an illustration, we set $\varepsilon=-10\tau_S$ and plot in Fig.~\ref{fig:consequences}d the exchange coupling $J/J_\infty$ as a function of $\Delta_{ST}/\tau_S$ and $\tau_T/\tau_S$. The exchange coupling is primarily suppressed by a large $\tau_T$, and this can only be alleviated by a strong opening of the singlet-triplet gap $\Delta_{ST}$. For example, if $\tau_T=\tau_S$ (which is the expected limit for many 1D devices, see Appendix~\ref{app:tcoupling}), $J$ is down by a factor 2 when $\Delta_{ST}=10\tau_S$. The exchange coupling only improves mildly up to $J=3J_\infty/4$ when $\Delta_{ST}=30\tau_S$ and to $J=5J_\infty/6$ when $\Delta_{ST}=50\tau_S$. Moreover, the exchange energy gets increasingly suppressed when detuning farther from the anti-crossing, which calls for even larger $\Delta_{ST}\gg|\varepsilon|$ to compensate for the finite $\tau_T$ \correction{(as further emphasized in section \ref{ref:subsubsop})}. It is, therefore, ultimately more efficient to design devices with a small $\tau_T/\tau_S$ ratio to limit exchange renormalization (which can be achieved in particular setups, as discussed in Appendix~\ref{app:tcoupling}). As an example, $J$ is reduced only by a factor 7/8 when $\tau_T=0.5\tau_S$ and $\Delta_{ST}=10\tau_S$. Although the suppression of $J$ may be accommodated by an enhancement of $\tau_S$ (which typically depends exponentially on the square root of the barrier height/tunnel gate bias), the renormalization is substantial at large detuning and must be at least accounted for in any realistic design of exchange-based multi-qubit systems.

\subsubsection{\correction{Symmetric operating point.}}\label{ref:subsubsop} A particularly important working point for exchange operations is the symmetric operating point (SOP) $\varepsilon\simeq-U$, which is protected against charge noise \citep{Reed:2016p110402, xue2021computing}. Near the SOP, the $(1,1)$ states are almost equally coupled to $(0,2)$ and $(2,0)$ states. To quantify the exchange coupling in the $(1,1)$ charge configuration, we therefore need to introduce both $(0,2)$ and $(2,0)$ states in the basis set (whereas Eq.~\eqref{eq:psbH} only involves the $(0,2)$ states). Again, we discard states that are only connected to $S(1,1)$ and $T(1,1)$ via second-order processes. The effective Hamiltonian in the minimal basis set $\{S(1,1),T(1,1),S(0,2),T(0,2),S(2,0),T(2,0)\}$ is then:
\begin{equation}
    H=\begin{pmatrix}
        0 & 0 & \tau_S & 0 & \tau_S & 0 \\
        0 & 0 & 0& \tau_T & 0 & \tau_T \\
        \tau_S & 0 & -\varepsilon'+U & 0 & 0 & 0 \\ 
        0 & \tau_T & 0 & -\varepsilon'+U_T^{(2)} & 0 & 0 \\
        \tau_S & 0 & 0 & 0 & \varepsilon'+U & 0 \\ 
        0& \tau_T & 0 & 0 & 0 & \varepsilon'+U_T^{(1)} 
    \end{pmatrix}\,
    \label{eq:Hst}
\end{equation}
where $U$ is the intra-dot charging energy (assumed identical in both dots), $U_T^{(1,2)}=U+\Delta_{ST}^{(1,2)}$, and $\Delta_{ST}^{(1,2)}$ is the singlet-triplet splitting in dots 1 and 2. The detuning $\varepsilon'$ is now measured with respect to the SOP. The lowest singlet and triplet eigenenergies of this Hamiltonian are plotted in Fig. \ref{fig:consequences}b, where the bending of the levels comes from the interaction with the $(0,2)$, $(2,0)$ states. After a Schrieffer-Wolff transformation, we obtain the exchange coupling as the splitting between the dressed $S(1,1)$ and $T(1,1)$ states:
\begin{equation}
    J^{\rm SOP}=J_\infty^{\rm SOP}-\frac{1}{2}\left(\frac{2\tau_T^2}{\varepsilon'+U_T^{(1)}}+\frac{2\tau_T^2}{-\varepsilon'+U_T^{(2)}}\right)\,,
   \label{eq:exchange}
\end{equation}
where we have introduced $J_\infty^{\rm SOP}=2\tau_S^2U/(U^2-\varepsilon'^2)$, the usual exchange coupling at the SOP when $\Delta_{ST}\rightarrow\infty$. Note that $\partial J_\infty^{\rm SOP}/\partial\varepsilon'$ is zero when $\varepsilon'=0$ (and that, in general, $\partial J^{\rm SOP}/\partial\varepsilon'$ has a zero near $\varepsilon'=0$): as stated above, the symmetric operating point is protected against charge noise on the detuning, and this protection can be enhanced to higher order \citep{abadillo2019enhancing}.

Yet, $\Delta_{ST}^{(1,2)}$ is generally much smaller than $U$, so that $U_T^{(1)}\simeq U_T^{(2)}\simeq U$. As a consequence, the exchange interaction at the SOP can be almost completely suppressed by a (remote) triplet with tunneling $\tau_T\simeq\tau_S$ and low $\Delta_{ST}$ as shown in Fig.~\ref{fig:consequences}b. Reaching a large enough $\Delta_{ST}^{(1,2)}\gg U$ to cancel out the triplet being impossible in most devices, the most practical way to limit exchange renormalization near the SOP is, therefore, to achieve $\tau_T\ll\tau_S$, as discussed in Appendix~\ref{app:tcoupling}. 

\section{Conclusions}\label{sec:conclusion}

In this work, we have characterized the properties of Wigner molecules in doubly occupied quantum dots as a function of the confinement and of the strength of Coulomb interactions. We have shown that the main fingerprint of the emergence of this correlated state is the strong suppression of the singlet-triplet splitting $\Delta_{ST}$. This suppression is caused by the separation of the particles along the weak confinement axis, and is in fact exponentially sensitive to the anisotropy of the dot. We have analyzed the spatial redistribution of the ground-state singlet and triplet densities in the molecules, and have discussed how they tend together to the classical limit at large interaction strength. In particular, we have quantified these effects for both electrons and holes in Si, Ge and GaAs, and find that Si is way more sensitive to Coulomb interactions than the other two materials. Indeed, the electrons and holes are heavier in Si, which suppresses the kinetic energy with respect to the Coulomb energy. Actually, nanowire-based hole spin qubits in silicon, also because of their anisotropic geometry, are most prone to Wigner molecularization. To limit these effects, the quantum dots must be designed as isotropic as possible. Complementarily, smaller dots show better resilience to Wigner molecularization than larger dots, even in anisotropic layouts.

The consequences of Wigner molecularization are not innocuous, as a low-lying triplet state can spoil Pauli spin blockade readout (PSB) and can reduce the exchange interactions between dots, thereby slowing down two-qubit gates. As for PSB, we have quantified the readout error, which turns out to be significant at small $\Delta_{ST}$, especially when the tunnel coupling $\tau_T$ between triplet states is comparable to the tunnel coupling $\tau_S$ between singlet states. Enhancing the confinement or reducing the dot anisotropy, either by tweaking the dot design or by tuning nearby gates, can increase $\Delta_{ST}$ and mitigate the error. Exchange interactions are also considerably suppressed when $\tau_T\simeq\tau_S$. The gap $\Delta_{ST}$ must then be much larger than the detuning $\varepsilon$ between the $(1,1)$ and $(0,2)$ singlets to alleviate the detrimental effects of the nearby triplet states. Such gaps may, however, be practically unreachable at large detuning. It may then be more efficient to design the dots so as to minimize $\tau_T/\tau_S$.

Our work emphasizes the necessity to consider the effects of Coulomb correlations on the properties of quantum dot devices currently explored for the realization of large scale multi-qubit systems. These effects may be more or less critical depending on the actual design of the quantum dots. They are not only relevant for the manipulation and readout of spin qubits~\cite{ercan2021strong,bosco2021squeezed,bosco2021fully}, but raise new opportunities and challenges, such as the detection of the fingerprints of correlations in the dynamics of quantum dots and quantum dot arrays, or when piling up additional electrons in one-dimensional channels~\cite{Shapir870}.

\section{Acknowledgements}

We acknowledge discussions with Silvano de Franceschi, Julia Meyer, Ekmel Ercan, Theodor Lundberg and Fernando Gonzalez-Zalba. This work was supported by the French national research agency (ANR project MAQSi).  

\appendix
\renewcommand\thefigure{\thesection.\arabic{figure}}
\setcounter{figure}{0}    

\section{Perturbation theory} 

\label{app:perturbation}

\begin{figure}
\centering
  \includegraphics[width=0.4\textwidth]{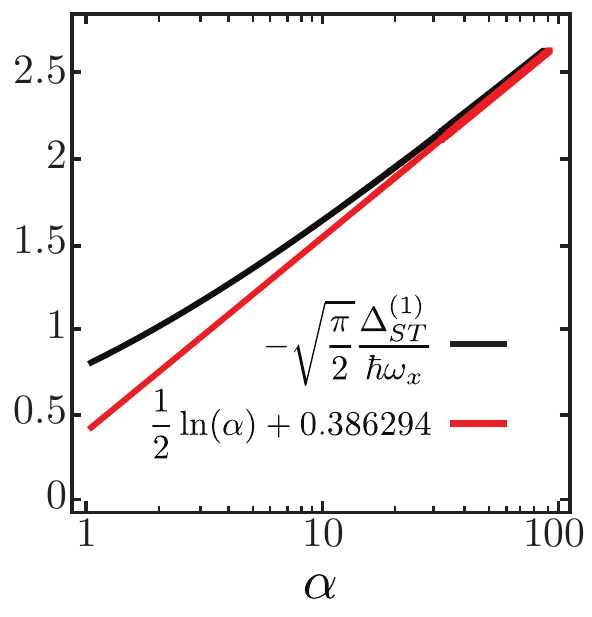}
\caption{Dependence of Eq.~\eqref{eq:DeltaST1} on the anisotropy $\alpha$, emphasizing the logarithmic divergence for $\alpha\gg1$.}
\label{fig:log}
\end{figure}

In this Appendix, we derive explicitly the leading correction in the interaction strength $\lambda_W$ to the singlet-triplet gap $\Delta_{ST}$ [Eq.~\eqref{eq:DeltaST1}]. Our starting point is the unitless Hamiltonian, Eq.~\eqref{eq:dimensionlessH}, in the anisotropic case $\alpha>1$. Coulomb interactions do not act on the spin degrees of freedom of the two-body wave functions, which nevertheless determine the even/odd behavior of the singlet/triplet states under the transformation $\mathbf r'\rightarrow -\mathbf r'$, describing particle exchange. As already discussed in the main text, in the non-interacting limit, the singlet and triplet solutions of Eq.~\eqref{eq:dimensionlessH} are $\varphi_S^{(0)}(x',y')=\phi_0(x')\phi_0(y')$ and $\varphi_T^{(0)}(x',y')=\phi_1(x')\phi_0(y')$, respectively, where $\phi_n(x')$ is the $n$-th eigenstate of the 1D harmonic oscillator. In the frame of the relative and normalized coordinates $\mathbf{r}'$ of Eq.~\eqref{eq:dimensionlessH}, these wave functions read:
\begin{subequations}
\begin{align}
    \varphi_S^{(0)}(x',y')&=\frac{\alpha^{1/4}}{\sqrt{2\pi}}\,e^{-\frac14\left(x'^2+\alpha y'^2\right)}\,,\\        \varphi_T^{(0)}(x',y')&=\frac{\alpha^{1/4}}{\sqrt{2\pi}}\,x'\,e^{-\frac14\left(x'^2+\alpha y'^2\right)}\,.
\end{align}
\end{subequations}
We remind that the leading correction to the singlet-triplet gap reads:
\begin{equation}
\begin{split}
\frac{\Delta_{ST}^{(1)}}{\hbar\omega_x}&=\lambda_W\left[\left\langle\varphi_T^{(0)}\left|\frac1{ r'}\right|\varphi_T^{(0)}\right\rangle-\left\langle\varphi_S^{(0)}\left|\frac1{ r'}\right|\varphi_S^{(0)}\right\rangle\right]\,,
 \end{split}
\end{equation}
which, in real-space representation, is given by the following two-dimensional integral: 
\begin{equation}
 \frac{\Delta_{ST}^{(1)}}{\hbar\omega_x}=\lambda_W\frac{\sqrt\alpha}{2\pi}\int dx'dy'\, \frac{x'^2-1}{r'}e^{-\frac12\left(x'^2+\alpha y'^2\right)}\,.
\end{equation}
After making the change of variables $\sqrt\alpha y'\rightarrow y'$ and switching to polar coordinates, the radial integration can be performed to find:
\begin{equation}\label{eq:deltaafterpolar}
 \frac{\Delta_{ST}^{(1)}}{\hbar\omega_x}=-\lambda_W\sqrt{\frac{2\alpha}\pi}\int_0^1 dt\sqrt{\frac{1-t^2}{1+(\alpha-1)t^2}}\,.
\end{equation}
Applying the algebraic identity:
\begin{multline}
    (\alpha-1)\sqrt{\frac{1-t^2}{1+(\alpha-1)t^2}}=\\
    \frac\alpha{\sqrt{[1-t^2][1+(\alpha-1)t^2]}}-\sqrt{\frac{1+(\alpha-1)t^2}{1-t^2}}\,,
\end{multline}
one can then identify in Eq.~\eqref{eq:deltaafterpolar} the standard expressions of the complete elliptic integrals of the first and second kind, namely~\footnote{Other conventions adopt the parametrization with $m=k^2$.}:
\begin{subequations}
\begin{align}
    \mbox{K}(m)&=\int_0^1dt\,\frac1{\sqrt{(1-t^2)(1-m\,t^2)}}\,,\\
    \mbox E(m)&=\int_0^1dt\,\sqrt{\frac{1-m\,t^2}{1-t^2}}\,.
\end{align}
\end{subequations}
We thus reach Eq.~\eqref{eq:DeltaST1} of the main text, whose dependence on $\alpha$ is plotted in Fig.~\ref{fig:log}. In particular, we highlight the logarithmic divergence of this expression in the strongly anisotropic limit $\alpha\gg1$, where:
\begin{multline}\label{eq:log}
    \frac{\sqrt\alpha}{\alpha-1}\left[\alpha\,\mbox K(1-\alpha)-\mbox E(1-\alpha)\right]\\\stackrel{\alpha\gg1}{\longrightarrow}\frac12\ln(\alpha)+0.386294\,.
\end{multline}

\section{Numerical solution of Hooke's atom} \label{app:model}
\setcounter{figure}{0}    
We solve Eq.~\eqref{eq:dimensionlessH} with the finite elements solver of Mathematica. This solver uses a triangular mesh and provides the first symmetric and anti-symmetric solutions $\ket{\varphi_S}$ and $\ket{\varphi_T}$ in a basis of linear interpolating functions, as well as their eigenenergies $E_S$ and $E_T$. To characterize the accuracy of the solution, we estimate the errors as:
\begin{subequations}
\begin{align}
{\rm Err}(S)&= \left|1-\frac{E_S}{\bra{\varphi_S}H'_{\mathbf{r}}\ket{\varphi_S}}\right|\,, \\
{\rm Err}(T)&= \left|1-\frac{E_T}{\bra{\varphi_T}H'_{\mathbf{r}}\ket{\varphi_T}}\right| \,.
\label{eq:errors}
\end{align}
\end{subequations}
The errors strongly depend on the problem parameters. For realistic confinement potentials, we find that the default Mathematica configuration yields errors in the $1-10\%$ range. To reduce the errors down to $0.01\%$ in the worst cases, we need to strongly refine the finite elements mesh by making the density of triangles up to 10$^4$ times larger near the Coulomb singularity.

\section{Quasi-3D extension} \label{app:model3d}

\setcounter{figure}{0}    
The model of Section~\ref{sec:model} assumes strict confinement in a 2D electron or hole gas. In practice, this gas has a non-zero thickness $t$, which reduces the strength of the Coulomb interactions. In an intermediate regime, where $t$ is not too large with respect to $\ell_x$ and $\ell_y$, finite thickness effects are merely expected to soften the formation of Wigner molecules and the renormalization of the singlet-triplet gap. However, Wigner molecularization must ultimately be suppressed at large enough $t$ because the particles regain freedom to cross each other without experiencing strong collisions.

In this appendix, we address finite thickness effects in the intermediate regime $0\ll t\ll\ell_x,\,\ell_y$. Therefore, we still assume strong enough vertical confinement so that the motion in the $(xy)$-plane can be separated from the motion along $z$. The latter can then be characterized by the ground-state wave function $\varphi_z(z)$, and can be integrated out of the equations of motion through the introduction of an effective in-plane Hamiltonian.

Generally, the particles are confined at a surface/interface in an heterostructure by a vertical electric field $F$. Assuming hard wall boundary conditions at $z=0$, $\varphi_z(z\ge 0)$ is then an Airy function, which is however practically difficult to integrate out. The ground-state Airy function can, nonetheless, be very well approximated by more easily integrable exponential or Gaussian functions in a variational approach. We find that the following trial wave function yields satisfactory results:
\begin{equation}
    \varphi_z(z,\beta)=\frac{2^{7/4}}{\pi^{1/4}}\beta^{3/4}z e^{-\beta z^2}\,.
\end{equation}
The variational parameter $\beta$ minimizes the ground-state energy of the vertical Hamiltonian
\begin{equation}
    H_z=\frac{p_z^2}{2m_\perp}+eFz\,,
\end{equation}
where $m_\perp$ is the confinement mass along $z$. The optimal ground-state wave function is therefore:
\begin{equation}
    \varphi_z(z,F)=\frac{4}{\sqrt{6\pi}}\frac{z}{\ell_z^{3/2}}\exp\left[-\frac{1}{(18\pi)^{1/3}}\frac{z^2}{\ell_z^2}\right]\,,
\end{equation}
where $\ell_z=[\hbar^2/(2m_\perp eF)]^{1/3}$ is the electrical confinement length. 

In the above assumptions, the total Hamiltonian of the system is $H_{\rm 3D}=H+H_z$, where $H$ is defined by Eq.~\eqref{eq:H}, and the two-particle wave function is $\psi(\mathbf{r}_1,\mathbf{r}_2)=\varphi_{xy}(x_1,y_1;x_2,y_2)\varphi_z(z_1,F)\varphi_z(z_2,F)$, where $\varphi_{xy}$ describes the in-plane motion. We can next integrate out the $z$ coordinate and define an effective Hamiltonian $H_{\rm 2D}\equiv\bra{\varphi_z(z_1,F)}\bra{\varphi_z(z_2,F)}H_{3D}\ket{\varphi_z(z_1,F)}\ket{\varphi_z(z_2,F)}$. The integral over the center of mass $z$ coordinate is performed analytically with Mathematica. The remaining integrals need, however, to be performed numerically in each finite element. This slows down the calculation by 1-2 orders of magnitude. 

\begin{figure}[h!]
\centering
  \includegraphics[width=0.4\textwidth]{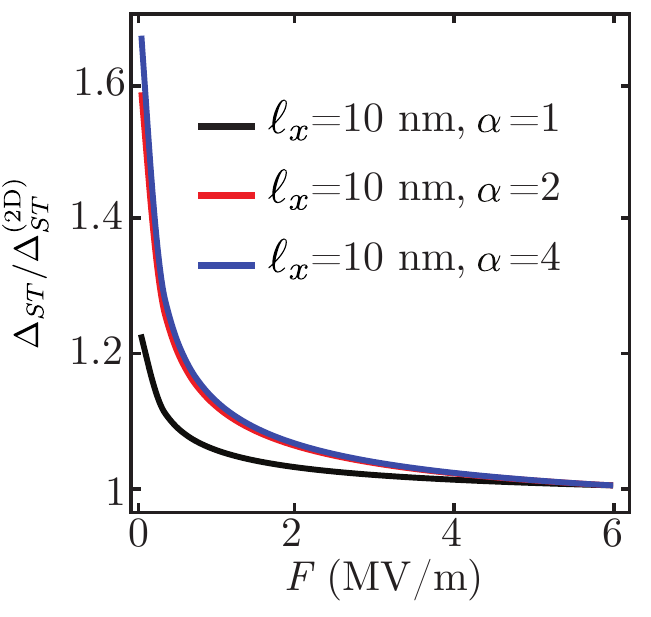}
\caption{Singlet-triplet splitting in the quasi-3D limit as a function of the vertical electric field $F$ in three different in-plane harmonic confinement potentials, for heavy-holes in Si. $\Delta_{ST}$ is normalized with respect to the splitting $\Delta_{ST}^{(2D)}$ in the strict 2D limit $F\to\infty$.}
\label{fig:w-fz}
\end{figure}

In Fig.~\ref{fig:w-fz}, we plot the singlet-triplet splitting $\Delta_{ST}$ as a function of the vertical electric field $F$ in three different in-plane harmonic confinement potentials, for heavy-holes in silicon ($m_\perp=m_0/(\gamma_1-2\gamma_2)=0.277\,m_0$). The results show that $\Delta_{ST}$ rapidly approaches the 2D limit $\Delta_{ST}^{(2D)}$ when increasing $F$. The difference remains $<20\%$ over the whole  $F>0.5$ MV/m range ($\ell_z<6.5$ nm). We emphasize, though, that the above approximations break down at very small $F$ when the vertical confinement energy becomes comparable to the in-plane harmonic confinement energy.

\section{Full configuration interaction}
\label{app:fullci}
\setcounter{figure}{0}    

The two-hole states in the realistic devices are computed with a full configuration interaction method. Namely, the two-particle wave function is expanded in the basis set of $N(N-1)/2$ Slater determinants built upon $N$ single particle wave functions $\psi_i(\mathbf{r})$ obtained with the 4 bands $\mathbf{k}\cdot\mathbf{p}$ model. Convergence of the singlet-triplet splitting is achieved using the $N=72$ lowest-lying hole states as input.

In the CI Hamiltonian, the matrix elements of the Coulomb interaction can be expressed as linear combinations of the Coulomb integrals:
\begin{equation}
U_{ijkl}=e^2\int d^3\mathbf{r}\int d^3\mathbf{r}^\prime\, \psi_i(\mathbf{r})\psi_j^*(\mathbf{r}) V(\mathbf{r},\mathbf{r}^\prime) \psi_k^*(\mathbf{r}^\prime)\psi_l(\mathbf{r}^\prime)\,,
\end{equation}
where $V(\mathbf{r},\mathbf{r}^\prime)$ is the Coulomb kernel (the potential created at point $\mathbf{r}^\prime$ by an unit charge at point $\mathbf{r}$). We can introduce the joint density:
\begin{equation}
\rho_{ij}(\mathbf{r})=\psi_i^*(\mathbf{r})\psi_j(\mathbf{r})\,.
\label{eq:rhoij}
\end{equation}
and the (complex) potential:
\begin{equation}
V_{kl}(\mathbf{r})=\int d^3\mathbf{r}^\prime\,V(\mathbf{r},\mathbf{r}^\prime)\rho_{kl}(\mathbf{r}^\prime)
\end{equation}
so that 
\begin{equation}
U_{ijkl}=e^2\int d^3\mathbf{r}\,\rho^*_{ij}(\mathbf{r})V_{kl}(\mathbf{r})\equiv\langle\rho_{ij}|V_{kl}\rangle
\end{equation}
is nothing else than the scalar product of $\rho_{ij}$ with $V_{kl}$. There are $N(N+1)/2$ independent $V_{kl}$'s to compute, which is the most time-consuming part of the CI, but can be parallelized trivially.

We use a finite volumes Poisson solver to calculate the $V_{kl}$'s. For that purpose, we expand the $\mathbf{k}\cdot\mathbf{p}$ wave functions as:
\begin{equation}
\psi_i(\mathbf{r})=\sum_\mu \varphi_{i\mu}(\mathbf{r})u_\mu(\mathbf{r})\,,
\end{equation}
where $\varphi_{i\mu}(\mathbf{r})$ is an envelope function (slowly varying at the atomic scale) and $u_\mu(\mathbf{r})$ is a bulk Bloch function (the topmost $|3/2,\pm 3/2\rangle$ and $|3/2,\pm 1/2\rangle$ valence band Bloch functions at $\Gamma$ for the 4 bands $\mathbf{k}\cdot\mathbf{p}$ model). The $u_\mu$'s are periodic over the underlying atomic lattice and are normalized so that:
\begin{equation}
\frac{1}{\Omega_0}\int_{\Omega_0} d^3\mathbf{r}\,u_\mu^*(\mathbf{r})u_\nu(\mathbf{r})=\delta_{\mu,\nu}\,,
\end{equation}
where $\Omega_0$ is the volume of a primitive unit cell. Terms $\mu\ne\nu$ in Eq.~\eqref{eq:rhoij} therefore integrate to zero on the scale of the unit cell and give rise to mostly short-range multipolar corrections to the Coulomb integrals that we neglect in the present work (an usual assumption in $\mathbf{k}\cdot\mathbf{p}$-based CI \cite{PhysRevB.104.035302,secchi2020inter}, essentially valid when the physics is dominated by long-range interactions, as is the case here). In this approximation,
\begin{equation}
\rho_{kl}(\mathbf{r})\approx\sum_\mu\varphi_{k\mu}^*(\mathbf{r})\varphi_{l\mu}(\mathbf{r})
\end{equation}
is slowly varying at the atomic scale. We can therefore compute $V_{kl}(\mathbf{r})$ as the solution of the macroscopic Poisson equation:
\begin{equation}
\varepsilon_0\mathbf{\nabla}\cdot\varepsilon_r(\mathbf{r})\mathbf{\nabla}V_{kl}(\mathbf{r})=-\rho_{kl}(\mathbf{r})\,,
\end{equation}
where $\varepsilon_r(\mathbf{r})$ is the material dependent dielectric constant in the device. This equation is practically solved using a finite volumes method on a real-space grid with appropriate boundary conditions (in particular, $V_{kl}(\mathbf{r})=0$ in the metals) \cite{VenitucciPRB2018}. We emphasize that all Coulomb integrals obtained that way are fully (and statically) screened by the dielectrics and metals.

\section{\correction{Wigner molecularization in the presence of valley  degrees of freedom}}
\label{app:wignervalleys}
\setcounter{figure}{0}    

\begin{figure}[t]
\centering
\includegraphics[width=0.4\textwidth]{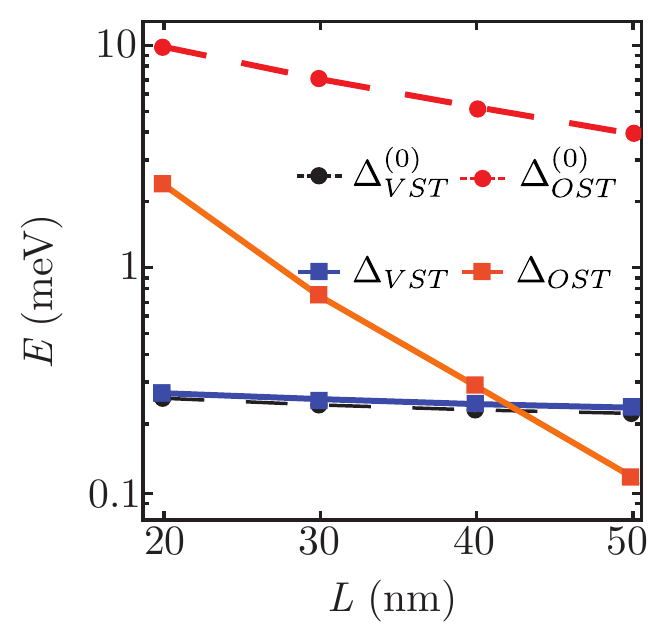}
\caption{\correction{Non-interacting (dashed lines) orbital splitting $\Delta_{OST}^{(0)}$, valley splitting $\Delta_{VST}^{(0)}$, and interacting (solid lines) orbital ($\Delta_{OST}$) and valley ($\Delta_{VST}$) singlet-triplet splittings as a function of the length $L$ of the gate in an electron device similar to Fig. \ref{fig:devices}a ($V_{\rm fg}=100$ mV, $V_{\rm bg}=-200$ mV on the substrate back gate).}}
\label{fig:deltaostvst}
\end{figure}

\correction{To illustrate the discussion of section \ref{subsec:materials} on the role of valleys in Wigner molecularization, we calculate the singlet-triplet splittings for electrons in 1D silicon devices identical to those of section \ref{subsec:devices}. For that purpose, we use an atomistic $sp^3d^5s^*$ tight-binding model \cite{niquet2009onsite} for the electronic structure of the dots \cite{BourdetPRB2018}, because it accounts for the valley-mixing and valley-orbit interactions missing in the simplest $\mathbf{k}\cdot\mathbf{p}$ approximations. The surface of the silicon channel is passivated by pseudo-hydrogen atoms that mimic a high bandgap material such as SiO$_2$. The CI calculations follow the same lines as for holes, with the atomic orbitals now playing the role of the Bloch function $u_\mu$.}

\correction{In the non-interacting system, the eigenstates can be sorted into ``orbital'' excitations (with different envelopes) and ``valley'' excitations (with similar envelopes but different valley composition). The non-interacting orbital splitting $\Delta_{OST}^{(0)}$ (which is by definition the non-interacting orbital singlet-triplet splitting) and the non-interacting valley splitting $\Delta_{VST}^{(0)}$ of the dot are plotted as a function of the length $L$ of the gate in Fig. \ref{fig:deltaostvst}. $\Delta_{OST}^{(0)}$ decreases with increasing $L$, as the lowest-lying orbital excitation is the $p_x$-like envelope oriented along the channel axis $x$. On the opposite, the much smaller valley splitting $\Delta_{VST}^{(0)}$ is almost constant, as it depends mainly on vertical confinement at the top and bottom interfaces of the channel \cite{Boykin:2004p115}. The orbital and valley excitations can then be followed adiabatically when switching on the Coulomb interactions to define the corresponding interacting orbital singlet-triplet splitting $\Delta_{OST}$ and interacting valley singlet-triplet splitting $\Delta_{VST}$. As expected from Section \ref{sec:model} and shown in Fig. \ref{fig:deltaostvst}, the orbital singlet-triplet splitting $\Delta_{OST}$ decreases even faster with increasing $L$ as the two electrons separate along $x$, while the valley singlet-triplet splitting $\Delta_{VST}$ is robust to Coulomb interactions. Ultimately, $\Delta_{OST}<\Delta_{VST}$ at large enough $L\gtrsim 43$ nm: the lowest-lying excitation becomes the orbital triplet.}

\correction{As a matter of fact, the orbital splitting is destabilized by the Coulomb interactions because mixing together the $s$-like ground-state with the first $p_x$-like orbital excitation localizes the electron on either side of the dot (and is hence the path to Wigner molecularization), whereas mixing together the two low-lying $s$-like valley states does not. Therefore, Wigner molecularization little interferes with valley physics (mostly through changes of the local interactions with the interfaces of the channel, see below).}

\correction{The conclusions of this work as to the effects of Wigner molecularization on exchange interactions and readout are thus about the same for electron as for hole quantum dots: a large valley splitting may anyway be superseded by a smaller orbital singlet-triplet splitting in a doubly occupied anisotropic quantum dot. This also raises the issue of the proper identification of the excitations (valley or orbital) in the spectroscopy of such quantum dots. Moreover, if the dots share the same valley states, and valley blockade is complete \cite{Hao:2014p3860}, $\tau_T\simeq 0$ for the valley triplet and the only relevant excitation for readout and exchange interactions is the orbital triplet, whether it is the lowest-lying triplet or not.}

\begin{figure}[h!]
\centering
\includegraphics[width=0.4\textwidth]{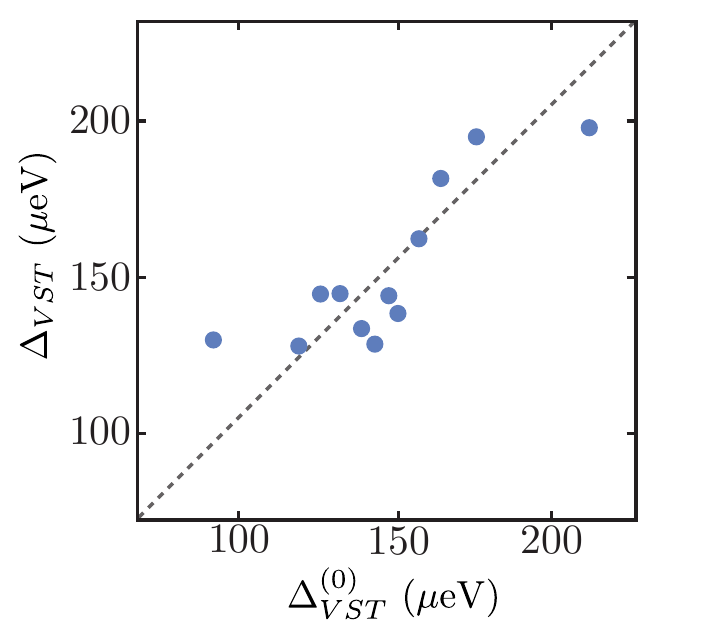}
\caption{\correction{Interacting valley singlet-triplet splitting  $\Delta_{VST}$ versus non-interacting valley splitting $\Delta_{VST}^{(0)}$ for different realizations of a surface roughness disorder ($L=30$ nm). The surface roughness profiles are characterized by a Gaussian auto-correlation function \cite{BourdetPRB2018} with rms $\Delta=0.4$ nm, and by a correlation length (length scale of the fluctuations) $\Lambda=1.5$ nm. The diagonal line is $\Delta_{VST}=\Delta_{VST}^{(0)}$.}}
\label{fig:deltavstSR}
\end{figure}

\correction{As it is well known, the non-interacting valley splitting is very sensitive to disorder (roughness) at the interfaces of silicon with the embedding materials (Ge, SiO$_2$, ...) \cite{Culcer:2010p205315,BourdetPRB2018,TariqPRB2019}. Roughness can, in particular, give rise to ``valley-orbit'' coupling between the orbital and valley degrees of freedom \cite{Friesen:2010p115324,Gamble:2013p035310, Abadillo-Uriel:2018p165438}. As shown in Fig. \ref{fig:deltavstSR}, the average $\Delta_{VST}^{(0)}$ is reduced by a factor $\simeq 2$ in a rough channel ($L=30$ nm) and shows significant device-to-device variability. The interacting $\Delta_{VST}$ follows the same trends, and is more affected by Coulomb interactions in some devices than in the pristine channel. This results from the relocalization of electrons by Wigner molecularization, which reshapes their interactions with the rough interfaces and the valley-orbit couplings \cite{ercan2021strong}. The interacting $\Delta_{VST}$ is not, however, systematically lower than the non-interacting $\Delta_{VST}^{(0)}$ in the present simulations (as is the case for the orbital splitting $\Delta_{OST}$). A better atomistic model for the Si/SiO$_2$ interface and for the inter-valley Coulomb scattering may nonetheless give more insights on the fine effects of Wigner molecularization on the valley splittings.}

\section{Tunnel coupling estimations}
\label{app:tcoupling}

\setcounter{figure}{0}    
\begin{figure}
\includegraphics[width=0.4\textwidth]{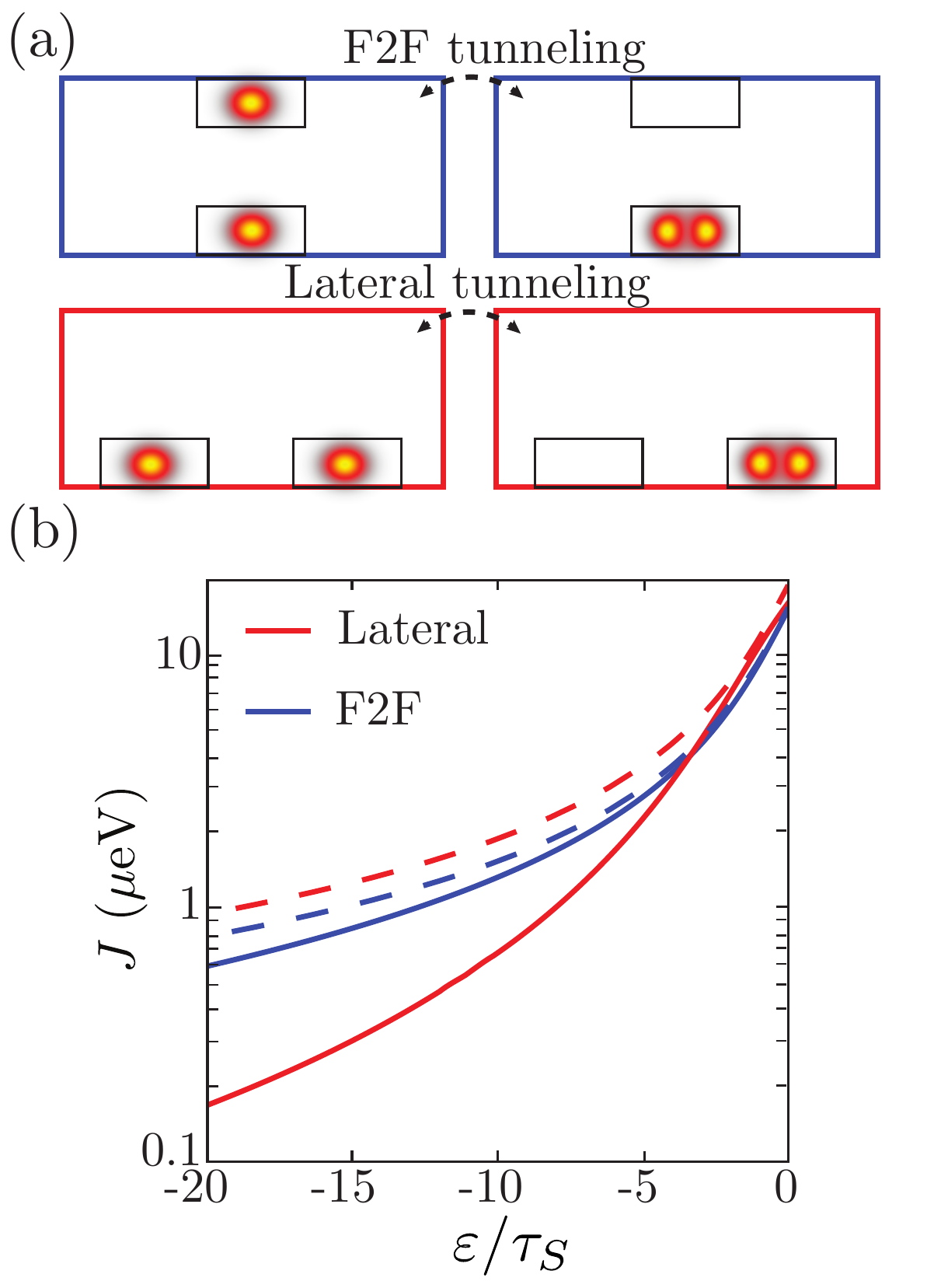}
\caption{Double-dot coupling and exchange interaction in 1D devices. (a) Schematics of the two considered configurations: face to face (F2F) and laterally coupled quantum dots. The gate lengths are $L=40$ nm and the separation between them is $S=40$ nm. (b) Exchange coupling $J$ as a function of detuning in F2F and lateral arrangements. The dashed lines are the expected value of $J_\infty$ obtained from the calculated tunnel couplings $\tau_S^{\rm F2F}=15.5$ $\mu$eV and $\tau_S^{\rm Lat}=19.1$ $\mu$eV). The solid lines account for triplet tunneling [Eq.~\eqref{eq:J02}]. The average front gate is $\overline{V}_{\rm fg}=-55$ mV. $\Delta_{ST}\approx 180\ \mu$eV is almost the same in both arrangements; $\tau_T/\tau_S\approx 1$ in the lateral case, while $\tau_T/\tau_S\approx 0$ in the F2F case.}
\label{fig:tratios}
\end{figure}

In Section~\ref{sec:implications}, we have shown that a low-lying triplet state can significantly suppress the exchange interaction between neighboring spins. This suppression is ruled by the ratio $\tau_T/\tau_S$ between the triplet and singlet tunnel couplings. In this appendix, we perform simulations of realistic double quantum dot devices to quantify this ratio and the exchange interactions.

We consider holes in a double quantum dot embedded within a 1D silicon channel similar to  Fig.~\ref{fig:devices}a. The dots can be arranged in two different configurations (see Fig.~\ref{fig:tratios}): they can either be coupled laterally along the channel or coupled face-to-face (F2F) across the channel. The length of the gates is $L=40$ nm, and the dots are separated by $S=40$ nm in both cases (the widths of the F2F and laterally coupled devices are $W_{\rm F2F}=80$ nm and $W_{\rm Lat}=35$ nm, and the gates overlap the channel by $20$ nm). In the lateral arrangement, the holes tunnel along the weak confinement axis, whereas in the F2F arrangement, they tunnel perpendicular to that axis. There is no exchange gate controlling the overall strength of tunneling in these devices (which are only used to illustrate the general trends). The tunnel couplings are extracted from full CI calculations of the anti-crossings between the $(1,1)$ and $(0,2)$ states.

We have evaluated the tunnel couplings $\tau_T$ and $\tau_S$ as a function of the average front gate voltage of the two dots, $\overline{V}_{\rm fg}=(V_{\rm fg}^{(1)}+V_{\rm fg}^{(2)})/2$, for both laterally coupled and F2F arrangements. In laterally coupled dots, we find a constant ratio $\tau_T/\tau_S\approx 1$. The quantum dots are indeed pretty anisotropic (1D-like), which makes the singlet and triplet wave functions in dot 2 look alike seen from dot 1 on the side. In the F2F configuration, we find on the opposite that $\tau_T/\tau_S$ remains negligible over the whole bias voltage range. This can be expected from the symmetry of the $T(1,1)$ and $T(0,2)$ states: the wave function of the $T(0,2)$ state changes sign when the two particles are mirrored with respect to the $(yz)$ symmetry plane perpendicular to the channel, while the $T(1,1)$ state remains invariant. As a consequence, tunneling between the triplet states is forbidden in the F2F configuration.

We finally analyze the impact of $\tau_T$ on the exchange coupling in each configuration. The exchange energy $J$ is plotted in Fig.~\ref{fig:tratios}b as a function of the energy detuning $\varepsilon$ between the dots, for an average $\overline{V}_{\rm fg}=-55$ mV. The trends are very different in the lateral and F2F arrangements, even though the lowest-lying $(0,2)$ triplet lies at the same $\Delta_{ST}\approx 180$ $\mu$eV in both cases. As expected, for F2F-coupled dots, the exchange interaction is not much affected by the lowest-lying triplet state given that $\tau_T/\tau_S\approx 0$. The weak renormalization of the exchange coupling likely results from to the interactions with higher-lying triplets. On the other hand, the laterally coupled dots display much lower exchange coupling since $\tau_T/\tau_S\approx 1$, which induces a strong bending of the triplet states energy. 
To conclude, $\tau_T/\tau_S$ is in general expected to be much larger when the particles tunnel along the weak confinement axis, and to reach unity in very anisotropic dots. In devices taking advantage of the anisotropy of the dots to enhance Rabi frequencies for example~\cite{MichalPRB2021,bosco2021squeezed,bosco2021fully}, the orientation of the weak confinement axis must, therefore, be carefully chosen in order not to hinder PSB readout and exchange interactions.

\bibliography{bib}

\end{document}